\documentclass[article,9pt]{IEEEtran}

\usepackage[]{graphicx}    
\newcommand{\ignore}[1]{}  
\usepackage{color}
\usepackage{cite}
\usepackage{epsfig,psfrag}
\usepackage{amssymb,amsmath,amsfonts}
\usepackage{bm}
\usepackage{tabularx}
\usepackage{multirow}
\usepackage{subcaption}
\usepackage{setspace}
\usepackage{amsthm}
\usepackage{tikz}
\usepackage{pst-node}
\usepackage{acronym}
\usepackage[yyyymmdd,hhmmss]{datetime}
\usepackage{notation}
\usetikzlibrary{arrows,backgrounds,calc,positioning,shapes,shadows}

\providecommand{\ist}{\hspace*{.3mm}}
\providecommand{\rmv}{\hspace*{-.3mm}}

\providecommand{\nn}{\nonumber}
\newcommand{\T}{\mathrm{T}}

\newcolumntype{L}[1]{>{\raggedright\arraybackslash}p{#1}}
\newcolumntype{C}[1]{>{\centering\arraybackslash}p{#1}}
\newcolumntype{R}[1]{>{\raggedleft\arraybackslash}p{#1}}

\acrodef{iqr}[IQR]{interquartile range}
\acrodef{gmm}[GMM]{Gaussian mixture model}
\acrodef{da}[DA]{data association}
\acrodef{mmse}[MMSE]{minimum mean-square error}
\acrodef{ps}[PS]{Potential Source}
\acrodef{po}[PO]{potential object}
\acrodef{pmf}[pmf]{probability mass function}
\acrodef{iid}[iid]{independent and identically distributed}
\acrodef{rmse}[RMSE]{root-mean-squared error}
\acrodef{ospa}[OSPA]{optimal sub-pattern assignment}
\acrodef{bp}[BP]{belief propagation}
\acrodef{bpf}[BPF]{bootstrap particle filter}
\acrodef{upf}[UPF]{unscented particle filter}
\acrodef{pde}[PDE]{partial differential equation}
\acrodef{sde}[SDE]{stochastic differential equation}
\acrodef{ode}[ODE]{ordinary differential equation}
\acrodef{edh}[EDH]{exact Daum and Huang}
\acrodef{ledh}[LEDH]{localized exact Duam and Huang}
\acrodef{pfpf}[PFPF]{particle flow particle filter}
\acrodef{mcmc}[MCMC]{Markov Chain Monte Carlo}
\acrodef{smc}[SMC]{sequential Monte Carlo}
\acrodef{map}[MAP]{maximum a posteriori}
\acrodef{tdoa}[TDOA]{time-difference of arrival}
\acrodef{pfl}[PFL]{particle flow}
\acrodef{pda}[PDA]{probabilistic data association}
\acrodef{jpda}[JPDA]{Joint \ac{pda}}
\acrodef{phd}[PHD]{probability hypothesis density}
\acrodef{cphd}[CPHD]{cardinalized \ac{phd}}
\acrodef{mht}[MHT]{multi-hypothesis tracking}
\acrodef{slam}[SLAM]{simultaneous localization and mapping}
\acrodef{iid}[iid]{independent and identically distributed}
\acrodef{rfs}[RFS]{random finite sets}
\acrodef{ospa}[OSPA]{optimal sub-pattern assignment}
\acrodef{mospa}[MOSPA]{mean \ac{ospa}}
\acrodef{snr}[SNR]{signal-to-noise ratio}
\acrodef{roi}[ROI]{region of interest}

\definecolor{temporalgreen}{RGB}{0,128,0}
\definecolor{spatialred}{RGB}{255,0,0}
\definecolor{temporalblue}{RGB}{0,0,205}

\allowdisplaybreaks
\begin{document}
\title{Particle Flows for Source Localization in 3-D\\ Using TDOA Measurements}
\author{Wenyu Zhang, Mohammad Javad Khojasteh, and Florian Meyer\\[0mm]
University of California San Diego, La Jolla, CA\\[0mm]
Email: \{wez078, m1khojasteh, flmeyer\}@ucsd.edu
\vspace*{-5mm}
}

\maketitle

\begin{abstract} 
Localization using
time-difference of arrival (TDOA) has myriad applications, e.g., in passive surveillance systems and marine mammal research. In this paper, we present a Bayesian estimation method that can localize an unknown number of static sources in 3-D based on TDOA measurements. The proposed localization algorithm based on particle flow (PFL) can overcome the challenges related to the highly non-linear TDOA measurement model, the data association (DA) uncertainty, and the uncertainty in the number of sources to be localized. Different PFL strategies are compared within a unified belief propagation (BP) framework in a challenging multisensor source localization problem. In particular, we consider PFL-based approximation of beliefs based on one or multiple Gaussian kernels with parameters computed using deterministic and stochastic flow processes. Our numerical results demonstrate that the proposed method can correctly determine the number of sources and provide accurate location estimates. The stochastic flow demonstrates greater accuracy compared to the deterministic flow when using the same number of particles. 
\end{abstract}

\section{Introduction}
\label{sec:intro}

We consider the problem of localizing an unknown number of sources that emit unknown signal waveforms in a three-dimensional (3-D) scenario by passively recording their acoustic or radio signals.
Our approach relies on \ac{tdoa} measurements that can be extracted, e.g., from the cross-correlation function of multiple spatially separated receivers. \ac{tdoa}-based localization is particularly relevant in marine mammal research \cite{JanMeySny:J23} and passive surveillance systems \cite{quazi81,huang01,TesMeyBee:J20}.

\begin{figure}[ht!]
  \centering
  \includegraphics[scale=0.37]{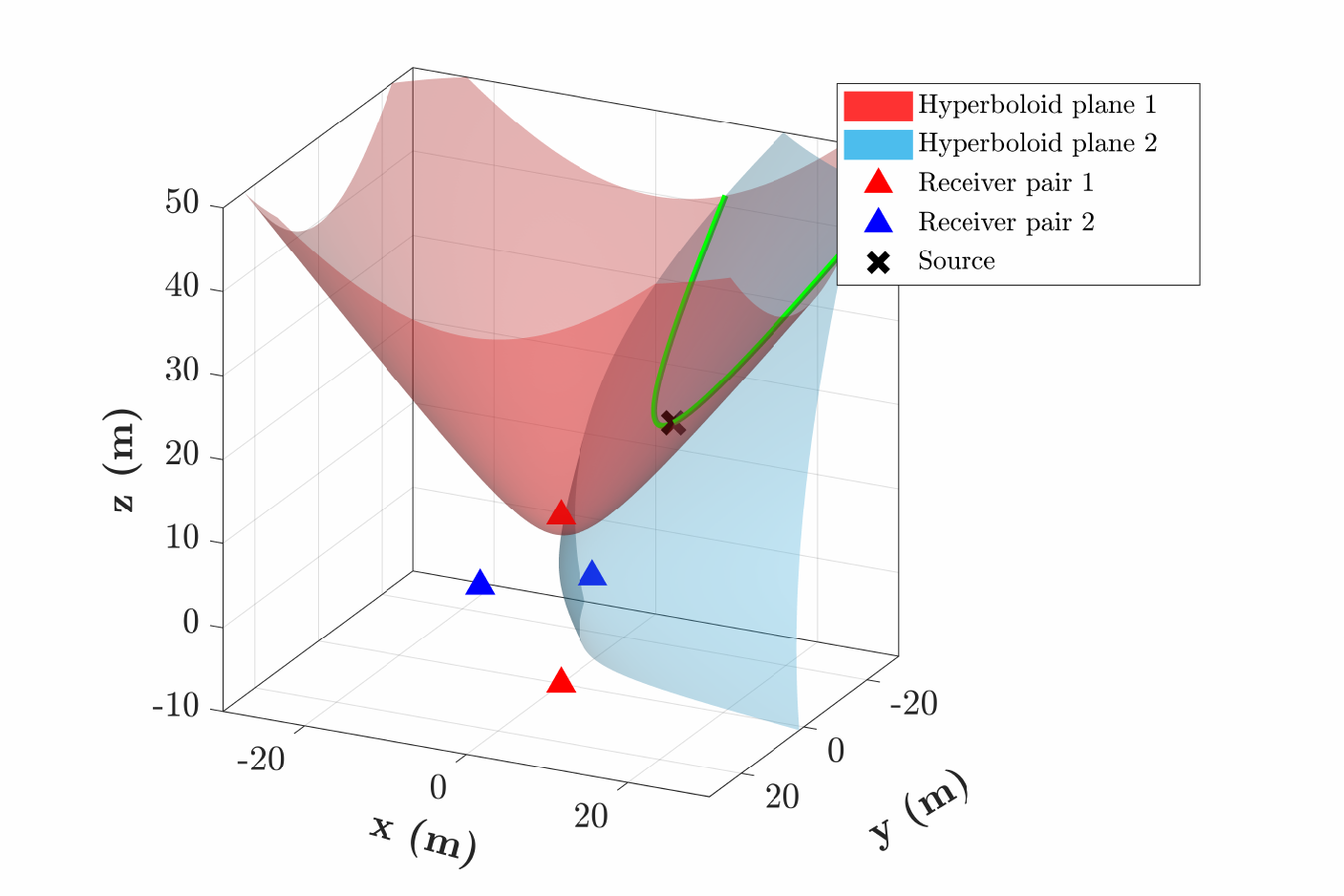}
  \caption{Source position and hyperboloids resulting from the TDOA measurements of two sensors. Each sensor consists of two receiver pairs. A dashed green line indicates the intersection of the two hyperboloids.}
  \label{fig:Hyperboloid}
  \vspace{-2mm}
\end{figure} 

\subsection{Source Localization Using \ac{tdoa} Measurements }
The TDOA measurement of a synchronized receiver pair will lead to 3-D location information on a hyperboloid \cite{ZhaMey:J24}. The hyperboloids related to a single source provided by multiple receiver pairs are expected to overlap closely with the actual location of the source.  Fig. \ref{fig:Hyperboloid} demonstrates the intersected curve (illustrated with the dashed green line) of two hyperboloids defined by 3-D \ac{tdoa} measurements of two receiver pairs. The localization of multiple sources is more challenging due to \ac{da} uncertainty \cite{MeyKroWilLauHlaBraWin:J18}, i.e., multiple TDOA measurements are generated by each receiver pair, but it is unknown which measurement was originated by which source. Furthermore, if \ac{snr} is low and in the presence of fading in the propagation paths, the signal of certain sources might be missing at some receivers, and false \ac{tdoa} measurements not related to a source might be erroneously extracted, which leads to missed detections and false alarms, respectively. Hence, the number of \ac{tdoa} measurements can differ among receiver pairs. In such a scenario, the number of sources to be localized is unknown, and the state space is high-dimensional and nonlinear. Previous works focused on the simple two-dimensional (2-D) setup \cite{MeyTesWin:C17} and single source localization \cite{gustafsson03tdoa}.

For the estimation of an unknown number of sources in the presence of DA uncertainty and nonlinear measurement models, particle-based belief propagation (BP) offers attractive solutions \cite{WilLau:J14,MeyBraWilHla:J17}. BP can outperform conventional probabilistic \ac{da} methods \cite{barShalom11}, where the number of sources is known, and \ac{mht} approaches \cite{reid79}, that typically rely on Gaussian approximations of posterior probability density functions (PDFs).

\subsection{Particle Degeneracy}
Particle-based BP for multisource localization typically relies on importance sampling \cite{AruMasGorCla:02} for particle-based computations \cite{MeyBraWilHla:J17}. Following the \ac{bpf} \cite{AruMasGorCla:02}, a prior or predicted PDF is typically used as a proposal PDF to draw samples that are then evaluated based on a likelihood function involving the measurements and their statistical model. However, the sampling efficiency of this strategy degrades significantly when the sources' state space has a higher dimension and source PDFs have complicated shapes \cite{BicLiBen:B08}. In particular, the sampling strategy of the \ac{bpf} fails in 3-D source localization problems where TDOAs are used as measurements \cite{ZhaMey:21,ZhaMey:J24}. This is because a very large number of particles is needed to ensure that a sufficient number of them fall into the region of high likelihood for measurement update. For a reasonable number of particles, the low number of particles that fall in the high likelihood region is insufficient to represent the underlying PDF accurately. This effect is typically referred to as  ``particle degeneracy''. To overcome particle degeneracy, the \ac{upf} \cite{MerDouFre:00} uses the unscented transform to obtain a Gaussian approximation of the posterior PDF to be used as a proposal PDF for sampling. On the other hand, \ac{pfl} \cite{DuaHua:07,DuaHua:09} involves migrating a random set of particles from a representation of a prior or predicted PDF to a representation of a posterior PDF density. Particle motion is described by an \ac{ode} or \ac{sde} established by using a log-homotopy function obtained from Bayes' rule in the Fokker–Planck equation \cite{DuaHua:07,DuaHua:09}. In certain applications, \ac{pfl}  can lead to strongly improved performance and reduced computational complexity compared to conventional particle-based techniques (see, e.g., \cite[Fig.~4]{MosChaCha:C16}). Different types of \ac{pfl} have been derived within a deterministic \cite{DuaHua:07,DuaHua:10} or stochastic \cite{DuaHua:13,DuaHuaNou:18} framework by solving the \ac{ode} and \ac{sde}, respectively. Stochastic flows involve a diffusion term with process noise added to particles in the flow. In some applications, their random nature can avoid numerical issues related to implementing the flow in discrete time. These numerical issues are sometimes referred to as stiffness \cite{DaiDau:22} of the flow. 

While \ac{pfl} methods display promising performance, there are generally no theoretical guarantees that an asymptotically optimal representation of the posterior PDF is obtained~\cite{BunGod:16}.
Certain \ac{pfl} can be used as a ``measurement-driven'' proposal PDF for importance sampling to perform asymptotically optimal estimation. 
Direct evaluation of the proposal PDF induced by \ac{pfl} is intractable, and various numerical methods for evaluating this proposal have been considered in the literature.
In particular, by focusing on the class of nonlinear Gaussian models, the work in~\cite{BunGod:16} provides numerical approximations of the proposal PDF, generated by \ac{pfl} within an importance sampling framework using Taylor series expansions. The method proposed in~\cite{BunGod:16} can be computationally demanding, and as an alternative, the authors of \cite{LiZhaoCoates:15} utilized auxiliary variables and filters to enhance the performance of importance sampling via \ac{pfl} more efficiently. The work in \cite{PalCoa:19} extended this framework to the stochastic Gromov's flow. Although the auxiliary methods \cite{LiZhaoCoates:15, PalCoa:19} are computationally less expensive than the method in \cite{BunGod:16}, they still rely on creating an auxiliary variable for each particle individually, which doubles the computational complexity of \ac{pfl}. Moreover, the choice of Gaussian covariance at each particle is problem-dependent and empirical. The work in~\cite{LiCoates:17} improves the estimation accuracy by creating an invertible mapping between the proposal and the prior PDF under the framework of the auxiliary particle filter. However, the invertible mapping is limited to deterministic flows such as \ac{edh} and \ac{ledh} as proposal PDF. \ac{pfl} has also been explored in the context of sequential Markov chain Monte Carlo methods for high-dimensional filtering~\cite{khan2017bayesian,li2017sequential}. 

\subsection{Contribution}

In this paper, we aim to solve a challenging 3-D multisource localization based on particle-based BP with \ac{da} uncertainty. Within our approach, every pair of receivers is considered a sensor that generates TDOA measurements, and the measurements provided by different sensors are processed sequentially. To address the particle degeneracy of conventional particle-based BP in high-dimensional space, we consider importance sampling based on deterministic and stochastic flow. As deterministic flows, we consider the \ac{edh} and the  \ac{ledh} \cite{DuaHua:07,DuaHua:10}. In addition, we investigate the stochastic so-called Gromov's flow \cite{DuaHuaNou:18}. 

The existing auxiliary-based methods \cite{LiZhaoCoates:15, PalCoa:19, LiCoates:17} can not be directly applied to the considered static source localization problem, as they rely on an underlying dynamic noise to distinguish the particle from the mean of the auxiliary PDF. We propose a \ac{gmm}-based method \cite{ZhaMey:21,ZhaMey:J24}, which is versatile for static and dynamic source problems. With fewer Gaussian kernels than the particles, the computational complexity can be greatly reduced since flow parameters only have to be computed for each Gaussian kernel. The covariance of each kernel is also analytically updated from the particles.
Our method can overcome the challenges related to the highly non-linear \ac{tdoa} measurement model, the \ac{da} uncertainty, and the uncertainty in the number of sources to be localized. 

Simulation results confirm that the number of sources can be determined correctly, and accurate location estimates can be obtained when the number of false alarms is low, and the probability of detection is high. The stochastic Gromov's flow achieves the best performance considering the localization accuracy and runtime, which exceeds existing state-of-the-art localization methods. Our simulation result promotes further research into stochastic flow since previous work shows a relatively small improvement in the accuracy of the state estimates \cite{PalCoa:19}.



\section{System Model}
\label{sec:measModel}

Consider the localization of multiple passive sources using $n_{\text{r}}$ receivers in 3-D in known locations $\V{q}^{(k)} = [q^{(k)}_{x} \ist q^{(k)}_{y} \ist q^{(k)}_{z}]^{\mathrm{T}}, k = 1,\dots,n_{\text{r}}$. The number of sources $n_{\text{t}}$ is unknown and their locations $\V{p}^{(j)} = [p_x^{(j)} \ist p_y^{(j)} \ist p_z^{(j)}]^{\mathrm{T}}, j = 1,\dots,n_{\text{t}}$ are random. Receivers can exchange their received signals and are perfectly synchronized, while the sources emit an unknown waveform at an unknown time.

\subsection{TDOA Measurements}
One effective method for acquiring location information of non-synchronized sources with unknown waveforms involves pairwise signal comparison at the receiver end.
In this way, for each receiver pair $(k,l)$, the signal of receiver $k$ and the signal of receiver $l$ are correlated, and time delays $z^{(m)}_{kl}$, $m = 1,\dots,n_{kl}$ related to peaks in the resulting cross-correlation function are extracted. These time delays are referred to as the TDOA measurements \cite{gustafsson03tdoa}. 
Each TDOA measurement $z^{(m)}_{kl}$ is related to a possible source location $\V{p}^{(j)}$ along a hyperboloid surface. For receiver pair $(k,l)$, the random TDOA $z^{(m)}_{kl}$ that was originated by source $j$ is modeled \vspace{0mm} as
\begin{align}
z^{(m)}_{kl} &= \frac{1}{c} \Big(  \| \V{p}^{(j)}-\V{q}^{(k)} \|  - \| \V{p}^{(j)} - \V{q}^{(l)} \| \Big) + v^{(m)}_{kl} \nn \\
&= h_{kl}(\V{p}^{(j)}) + v^{(m)}_{kl} \label{eq:tau} \\[-6mm]
\nn
\end{align}
where $c$ is the propagation speed of signal and $v^{(m)}_{kl}$ is the measurement noise which is zero-mean Gaussian with variance $\sigma_v^2$ and statistically independent across $m$ and across all $(k,l)$ pairs. Here, we consider that the source signal is received on a single and direct line of sight path. Modeling multipath propagation and refraction is subject to future work. The dependence of a measured TDOA $z^{(m)}_{kl}$ on the location $\V{p}^{(j)}$ of the generating source $j$ is described by the likelihood function $f(z^{(m)}_{kl}|\ist\V{p}^{(j)})$ that can be directly obtained from \eqref{eq:tau}.

\subsection{\Acp{ps}, \ac{da}, and Multisource Likelihood Function}

For simplicity of notation, we label each receiver pair $(k,l)$ by a single index $s = 1,\dots,S$, where $S$ is the total number of receiver pairs (a.k.a. sensors). We furthermore denote the two receivers of sensor $s$ by $(k_s,l_s)$ and the number of \ac{tdoa} measurements at sensor $s$ by $M_{s}\triangleq n_{k_s l_s}$. Finally,  we introduce $\V{z}_{s} \triangleq \big[z^{(1)\T}_{s} \cdots z^{(M_s)\T}_{s}\big]^{\T}\rmv\rmv\rmv$, i.e., the vector of TDOA measurements at sensor $s$.

With the assumption that one source can generate at most one measurement at a receiver pair and one measurement can originate from at most one source (a.k.a the ``the point source assumption'' \cite{BarWilTia:B11,Mah:B07,MeyKroWilLauHlaBraWin:J18}), we can sequentially processing measurements sensor by sensor and formulate the location of an unknown number of sources by introducing augmented \ac{ps} states $\V{y}_{s}^{(j)} \rmv\triangleq\rmv \big[\V{x}^{(j)\T}_{s} \ist r^{(j)}_{s} \big]^\T\rmv\rmv\rmv, j = 1,\dots,J_s$. The number of \acp{ps} $J_{s}$ at sensor $s$ is the maximum possible number of sources that have generated a measurement up to sensor $s$.

The augmented state of \ac{ps} $\V{y}_{s}^{(j)}$ incorporates an existence variable $r^{(j)}_{s} \rmv\in\rmv \{0,1\}$ along with \ac{ps}'s state variable $\V{x}^{(j)}_{s}\rmv\rmv$, which is the source position in our problem, i.e., $\V{x}^{(j)}_{s}\triangleq\V{p}^{(j)}$. $r^{(j)}_{s}$ models the existence/nonexistence of \ac{ps} $j$ in the sense that \ac{ps} $j$ exists at sensor $s$ if and only if $r^{(j)}_{s} \rmv\rmv=\rmv\rmv 1$. For nonexistent \acp{ps}, i.e., $r^{(j)}_{s} \!=\! 0$, the state $\V{x}^{(j)}_{s}$ is obviously irrelevant. Thus, all PDFs of nonexistent \ac{ps} states can be expressed as $f\big(\V{x}^{(j)}_{s}\rmv, {r}^{(j)}_{s} \!=\! 0 \big) = f^{(j)}_{s} f_{\text{D}}\big(\V{x}^{(j)}_{s}\big)$, where $f_{\text{D}}\big(\V{x}^{(j)}_{s}\big)$ is an arbitrary ``dummy PDF'' and $f^{(j)}_{s} \!\rmv\in [0,1]$ is a constant.
For sequential processing of sensors, a new \ac{ps} is introduced for each of the $M_s$ observed measurements at sensor $s$, and the total number of \acp{ps} is updated as $J_s = J_{s-1} + M_s$. All \acp{ps} that have been introduced at previous sensors are referred to as legacy \acp{ps}. There are $J_{s-1}$ legacy \acp{ps} and $M_s$ new \acp{ps}. To distinguish between legacy and new \acp{ps} at sensor $s$, we denote\vspace{-.5mm} by $\underline{\V{y}}^{(j)}_{s}, j = 1,\dots,J_{s-1}$ and by $\overline{\V{y}}^{(m)}_{s}, m = 1,\dots,M_s$ the augmented state of a legacy \ac{ps} and a new \ac{ps} states, respectively. 

The detection probability denoted as $p_{\text{d}}$ is assumed to be constant across sensors. The source represented by \ac{ps} $j \in \{1,\dots, J_{s}\}$ is detected (in the sense that it generates a measurement $z_s^{(m)}$) at sensor $s$ with probability $p_{\text{d}}$.  The statistical relationship of a measurement $z_s^{(m)}$ and a detected \ac{ps} state $\V{x}_s^{(j)}$ is described by the conditional PDF $f\big({z}^{(m)}_{s} | \V{x}^{(j)}_{s} \big)$, which is based on the measurement model of the sensor (c.f. \eqref{eq:tau} for the 3-D \ac{tdoa} in this work). In multisource localization, measurements are subject to \ac{da} uncertainty: it is unknown which measurement originated from which \ac{ps}, and a measurement may also be clutter, i.e., not originating from any \ac{ps}. Clutter measurements are modeled by a Poisson point process with mean $\mu_{\text{c}}$ and PDF $f_{\text{c}}\big( {z}^{(m)}_{s} \rmv\big)$. With the point source assumption, the association between $M_s$ measurements and $J_{s-1}$ legacy \acp{ps} at time $k$ can be modeled by an ``source-oriented'' \ac{da} vector $\V{a}_{s} = \big[a_{s}^{(1)} \cdots\ist a_{s}^{(J_{s-1})} \big]^{\T}\rmv\rmv$. The source-oriented association variable $a_{s}^{(j)}$ is $m \in \{1,\dots,M_s \}$ if \ac{ps} $j$ generates measurement $m$ and zero if \ac{ps} $j$ is missed by the sensor. For a legacy \ac{ps} $j$, the measurement model is represented by the factors 
\begin{align}
  q\big( \underline{\V{x}}^{(j)}_{s}\rmv\rmv\rmv,\rmv \underline{r}_s^{(j)}\rmv\rmv\rmv=\rmv1, a^{(j)}_{s}\rmv\rmv; {z}_{s} \rmv\big) \rmv\triangleq\rmv \begin{cases}
           \frac{p_{\mathrm{d}} f( {z}^{(m)}_s \rmv |\ist \underline{\V{x}}^{(j)}_s )} { \mu_{\mathrm{c}} f_{\mathrm{c}}( {z}^{(m)}_s )},\rmv\rmv\rmv\rmv\rmv & a^{(j)}_s \rmv\rmv=\rmv\rmv m \rmv\rmv\in\rmv\rmv \{1,\dots,M_s \}  \\[1mm]
          1 - p_{\mathrm{d}},\rmv\rmv\rmv\rmv\rmv& a^{(j)}_s \rmv\rmv=\rmv\rmv 0. \ist
         \end{cases}
         \nn\\[-4.0mm]
          \label{eq:singleLikelihood1}
\end{align}
and $ q\big( \underline{\V{x}}^{(j)}_{s}\rmv\rmv\rmv,\rmv \underline{r}_s^{(j)}\rmv\rmv\rmv=\rmv0, a^{(j)}_{s}\rmv\rmv; \V{z}_{s} \rmv\big)\rmv\triangleq\rmv 1(a^{(j)}_{s})$, where $1(a)$ denotes the indicator function of the event $a \rmv=\rmv 0$ (i.e., $1(a) \rmv=\rmv 1$ if $a \rmv=\rmv 0$ and $0$ otherwise) \cite{MeyKroWilLauHlaBraWin:J18}. We also introduce the ``measurement-oriented'' \ac{da} vector $\V{b}_{s} = \big[b_{s}^{(1)} \cdots\ist b_{s}^{(M_s)} \big]^{\T}\rmv\rmv$ to obtain a scalable and efficient message passing algorithm (see \cite{WilLau:J14,MeyKroWilLauHlaBraWin:J18,MeyWil:J21} for details). The measurement-oriented association variable $b_{s}^{(m)}$ is $j \in \{1,\dots,J_{s-1} \}$ if measurement $m$ originated from legacy \ac{ps} $j$ and zero if it originated from clutter or a newly detected \ac{ps}. Furthermore, for a new \ac{ps}, the measurements model is represented by the \vspace{.5mm}factors
\begin{align}
& v\big( \overline{\V{x}}^{(m)}_{s}\rmv\rmv\rmv, \overline{r}^{(m)}_{s}\rmv\rmv\rmv=1, b^{(m)}_{s}\rmv; {z}_{s}^{(m)} \big)  \nn \\[.5mm]
&\hspace{2mm} \triangleq \begin{cases}
     0,  & \hspace{-1mm} b^{(m)}_{s} \rmv\rmv\in\rmv\rmv  \{ 1,\dots,J_{s-1} \}\\[2mm]
     { \ist \frac{f\big({z}_s^{(m)}  \big| \overline{\V{x}}^{(m)}_{s} \big)}{ \mu_{\text{c}} f_{\text{c}}\big( {z}_{s}^{(m)} \big)} }\mu_{\text{b}} \ist f_{\text{b}}\big(\overline{\V{x}}^{(m)}_{s}\big), & \hspace{-1mm} b^{(m)}_{s} \rmv\rmv=\rmv\rmv 0 
  \end{cases} \hspace{1.8cm}
  \nn\\[-3mm]
  \label{eq:singleLikelihood2}
\end{align}
and $v\big( \overline{\V{x}}^{(m)}_{s}\rmv\rmv\rmv, \overline{r}^{(m)}_{s}\rmv\rmv\rmv=0, b^{(m)}_{s}\rmv; {z}_{s}^{(m)} \big)  \rmv\triangleq\rmv f_{\text{D}}\big(\overline{\V{x}}^{(m)}_{s}\big)$ \cite{MeyKroWilLauHlaBraWin:J18}. In  \eqref{eq:singleLikelihood2}, the birth of new \acp{ps} is modeled by a Poisson point process with mean $\mu_{\text{b}}$ and PDF $f_{\text{b}}\big( \overline{\V{x}}_{s} \big)$. 

\subsection{Joint Posterior PDF and Problem Formulation}
With the previous single \ac{ps} likelihood, we can obtain the joint posterior PDF of all random variables given the sensors' measurements. In particular, by using common assumptions \cite{MeyKroWilLauHlaBraWin:J18}, we obtain the following expression for the joint posterior PDF, i.e., 
\begin{align}
  &f\big( \V{y}_{0:s}, \V{a}_{1:s}, \V{b}_{1:s} \big| \V{z}_{1:s} \big) \nn \\[0.7mm]
  &\hspace{0mm} \propto  \Bigg(\prod^{J_{0}}_{j''=1} f\big(\V{y}^{(j'')}_{0} \big)  \Bigg)  \prod^{s}_{s'=1}   \Bigg(\prod^{J_{s'-1}}_{j'=1} f\big(\underline{\V{y}}^{(j')}_{s'}\big|\V{y}^{(j')}_{s'-1}\big)  \Bigg)   \nn\\[0.5mm] 
  &\hspace{4.5mm}\times \Bigg( \prod^{J_{s'-1}}_{j=1}  q\big( \underline{\V{x}}^{(j)}_{s'}\!, \underline{r}^{(j)}_{s'}\!\rmv, a^{(j)}_{s'}\rmv; \V{z}_{s'} \big)\rmv\rmv \prod^{M_{s'}}_{m'=1} \Psi_{j\rmv,m'}\big(a_{s'}^{(j)}\rmv,b_{s'}^{(m')}\big) \rmv\Bigg)  \nn\\[1.5mm]
  &\hspace{4.5mm}\times  \hspace{1mm} \prod^{M_{s'}}_{m=1} \rmv v\big( \overline{\V{x}}^{(m)}_{s'}\!, \overline{r}^{(m)}_{s'}\!, b^{(m)}_{s'}\rmv; {z}_{s'}^{(m)} \big)\hspace{-.1mm}.
  \label{eq:jointPosteriorComplete} \\[-4.5mm]
  \nn
\end{align}
Here, the $f\big( \V{y}^{(j)}_{0}\big)$ are uninformative prior PDF of \ac{ps} $j$ before any measurement update. These prior PDFs are typically uniformly distributed on a certain \ac{roi}. The binary indicator function $\Psi_{j\rmv,m}\big(a_{s}^{(j)}\rmv,b_{s}^{(m)}\big)$ in \eqref{eq:jointPosteriorComplete} checks association consistency of a pair of source-oriented and measurement-oriented variables $\big(a_{s}^{(j)}\rmv,b_{s}^{(m)}\big)$ in that $\Psi_{j\rmv,m}\big(a_{s}^{(j)}\rmv,b_{s}^{(m)}\big)$ is zero if 
$a_{s}^{(j)} \rmv= m, b^{(m)}_{s} \rmv\neq\rmv j$ or $b^{(m)}_{s} \rmv=\rmv j, a_{s}^{(j)} \rmv\neq\rmv m$ and one otherwise (see \cite{WilLau:J14,MeyKroWilLauHlaBraWin:J18} for details). The state transition PDF $f\big(\underline{\V{y}}^{(j)}_{s}\big|\V{y}^{(j)}_{s-1}\big)$ between two sensors is a $\delta$-function $\delta(\underline{\V{y}}^{(j)}_{s}\rmv\rmv-\V{y}^{(j)}_{s-1})$.

With the joint posterior PDF presented in \eqref{eq:jointPosteriorComplete}, we formulate the detection and localization problem we want to solve. We consider the problem of simultaneous detection and localization of an unknown number of static sources based on all measurements $\V{z}_{1:S}$ collected by a total number of $S$ multiple receiver pairs (sensors). Source detection is performed by comparing the existence probability $p\big(r_{S}^{(j)}\! \rmv=\rmv 1 \big| \V{z}_{1:S} \big)$ with a threshold $P_{\text{th}}$, i.e., \ac{ps} $j \in \{1,\dots,J_S\}$ is declared to exist if $p\big(r_{S}^{(j)}\! \rmv=\rmv 1 \big| \V{z}_{1:S} \big) \rmv>\rmv P_{\text{th}}$. Note that $p\big(r_{S}^{(j)}\! \rmv=\rmv 1 \big| \V{z}_{1:S} \big) \rmv= \int f\big(\V{x}_S^{(j)}, r_S^{(j)}\! \rmv=\rmv 1 \big| \V{z}_{1:S}\big) \ist\mathrm{d}\V{x}_S^{(j)}$.
For existent \acp{ps}, state estimation is performed by calculating the \ac{mmse} estimate \cite{Poo:B94}  as
\vspace{0mm}
\begin{equation}
\hat{\V{x}}_S^{(j)}
\ist\triangleq \int \V{x}_S^{(j)} f\big(\V{x}_S^{(j)} \big| r_S^{(j)} \rmv=\rmv 1, \V{z}_{1:S}\big) \ist\mathrm{d}\V{x}_S^{(j)}
\label{eq:mmseEst}
\end{equation} 
where $f\big(\V{x}_S^{(j)} \big| r_S^{(j)} \rmv=\rmv 1, \V{z}_{1:S}\big) \rmv=\rmv f\big(\V{x}_S^{(j)}, r_S^{(j)} \rmv=\rmv 1 \big| \V{z}_{1:S}\big)/ p\big(r_{S}^{(j)}$ $=\rmv 1\big| \V{z}_{1:S} \big)$.

Both source detection and estimation require the marginal posterior PDF $f\big(\V{x}_S^{(j)}, r_S^{(j)} \big| \V{z}_{1:S})\rmv \triangleq f\big(\V{y}_S^{(j)} \big| \V{z}_{1:S})$, $j \in \{1,\dots,J_S\}$. Using the Markovian property of the joint posterior PDF, we can update the marginal posterior PDF $f\big(\V{x}_s^{(j)}, r_s^{(j)} \big| \V{z}_{1:s})$ by processing TDOA measurements $\V{z}_{s}$ sequentially across sensors $s \in \{1,\dots,S\}$.

\subsection{Message Passing Operations}
\begin{figure}[t!]
\centering

\hspace{1mm}\includegraphics[scale=1]{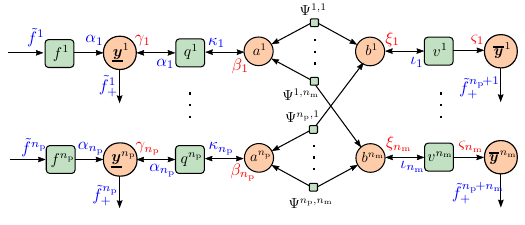}

\caption{Factor graph for message passing of an unknown number of sources at a single sensor $s$, corresponding to the propagation of the joint PDF \eqref{eq:jointPosteriorComplete}. Messages calculated using \ac{pfl} are depicted in \textcolor{spatialred}{red}. These messages are calculated based on messages depicted in \textcolor{blue}{blue}. The following short notations are used: 
$n_{\mathrm{m}} \rmv\triangleq M_{s}$, 
$n_{\mathrm{p}} \rmv\triangleq J_{s-1}$,
$\underline{\V{y}}^{j} \rmv\triangleq \underline{\V{y}}_{s}^{(j)}$,
$\overline{\V{y}}^{\ist m} \rmv\triangleq \overline{\V{y}}_{s}^{(m)}\rmv$,
$\V{a}^j \rmv\triangleq \V{a}^{(j)}_s$,
$\V{b}^m \rmv\triangleq \V{b}^{(m)}_s$,
\textcolor{temporalgreen}{$f^j \rmv\triangleq f\big(\underline{\V{y}}^{(j)}_{s} \big| \V{y}^{(j)}_{s-1}\big)$},
\textcolor{temporalgreen}{$q^j \rmv\triangleq q\big( \underline{\V{x}}^{(j)}_{s}\rmv, \underline{r}^{(j)}_{s}\rmv, a_{s}^{(j)};\V{z}_{s} \big)$}, 
\textcolor{temporalgreen}{$v^m \rmv\triangleq v\big( \overline{\V{x}}^{(m)}_{s} \rmv,\overline{r}^{(m)}_{s} \rmv,b_{s}^{(m)};\V{z}^{(m)}_{s} \big)$},
\textcolor{temporalgreen}{$\Psi^{j,m} \rmv\triangleq \Psi_{j,m}\big(a^{(j)}_{s} \rmv,b^{(m)}_{s}\big)$}, 
\textcolor{red}{$\gamma_j \triangleq \gamma_{s}^{(j)}\big(\underline{\V{y}}^{(j)}_{s} \big)$}, 
\textcolor{red}{$\beta_j \triangleq \beta^{(j)}_{s}\big(a^{(j)}_{s}\big)$},  
\textcolor{red}{$\xi_m \triangleq \xi^{(m)}_{s}\big(b^{(m)}_{s}\big)$}, 
\textcolor{red}{$\varsigma_{m} \triangleq \varsigma^{(m)}_s\big(\overline{\V{y}}^{(m)}_{s} \big)$}, 
\textcolor{blue}{$\tilde{f}^j\rmv\triangleq \tilde{f}\big(\V{y}_{s-1}^{(j)}\big)$}, 
\textcolor{blue}{$\alpha_j \triangleq \alpha_s\big(\underline{\V{y}}^{(j)}_{s}\big)$}, 
\textcolor{blue}{$\kappa_j \triangleq \kappa^{(j)}_{s}\big(a^{(j)}_{s} \big)$}, 
\textcolor{blue}{$\iota_{m} \triangleq \iota_s^{(m)}\big(b^{(m)}_{s} \big)$}, and 
\textcolor{blue}{$\tilde{f}^j_+\rmv\triangleq \tilde{f}\big(\V{y}_{s}^{(j)}\big)$}.
}
\label{fig:factorGraphU}
\end{figure}

Typically, calculating $f\big(\V{x}_s^{(j)}, r_s^{(j)} \big| \V{z}_{1:s}), s = 1,\dots,S$ by directly marginalization over \eqref{eq:jointPosteriorComplete} is infeasible due to a large number of parameters. As in \cite{MeyBraWilHla:J17,MeyKroWilLauHlaBraWin:J18}, we approximate the marginal posterior PDF by performing a loopy BP on the factor graph in Fig. \ref{fig:factorGraphU} and passing messages forward across sensor indexes. More specifically, let's consider the message passing for a single sensor $s$. The belief computed for sensor $s-1$, i.e., $\tilde{f}\big(\V{x}_{s-1}^{(j)}, r_{s-1}^{(j)} \big)\approx f\big(\V{x}_{s-1}^{(j)}, r_{s-1}^{(j)} \big| \V{z}_{1:s-1})$, is directly used as prior information for computations at sensor $s$ since a Dirac delta function describes state-transition between sensor indexes. For computations related to sensor $s$, this ``prior belief'' is denoted as  $\alpha_s\big(\underline{\V{y}}^{(j)}_{s}\big)$. For legacy \acp{ps}, the messages $\beta_s^{(j)}\big(a_s^{(j)}\big)$ passed from factor nodes $q\big( \underline{\V{x}}^{(j)}_{s}\!, \underline{r}^{(j)}_{s}\!, a^{(j)}_{s}\rmv; \V{z}_{s} \big)$ to variable nodes $a^{(j)}_{s}$ are calculated as
\begin{align}
  &\beta_s^{(j)}\big(a_s^{(j)}\big) = \rmv\int\rmv q\big( \underline{\V{x}}^{(j)}_{s}\!, 1, a^{(j)}_{s}\rmv; \V{z}_{s} \big) \alpha_s^{(j)}\!\big(\underline{\V{x}}_s^{(j)}\!,1\big) \mathrm{d}\underline{\V{x}}_s^{(j)} \nn\\
  &\hspace{52.5mm}+ 1\big(a_s^{(j)}\big)\alpha_{\mathrm{n},s}^{(j)} \ist.
  \label{eq:messageBeta}
\end{align}
where $\alpha_{\mathrm{n},s}^{(j)} \rmv=\rmv \int \alpha_s^{(j)}\big(\underline{\V{x}}_s^{(j)}\rmv\rmv,\underline{r}_s^{(j)} \rmv=\rmv0\big) \mathrm{d} \underline{\V{x}}_s^{(j)}$. For new \acp{ps}, messages $\xi_k^{(m)}\big(b_k^{(m)}\big)$ are calculated similarly (see \cite[Section~IX]{MeyKroWilLauHlaBraWin:J18}).
Now, probabilistic \ac{da} is performed by using the iterative BP-based algorithm \cite[Section~IX-A3]{MeyKroWilLauHlaBraWin:J18} with input messages $\beta_s^{(j)}\big(a_s^{(j)}\big)$, $j \in \{1,\dots,J_{s-1}\}$ and $\xi_s^{(m)}\big(b_s^{(m)}\big)$, $m \in \{1,\dots,M_s\}$. After convergence, corresponding output messages $\kappa_s^{(j)}\big(a_s^{(j)}\big)$, $j \in \{1,\dots,J_{s-1}\}$ and $\iota_s^{(m)}\big(b^{(m)}_{s} \big)$, $m \in \{1,\dots,M_s\}$ are available for legacy \acp{ps} and new \acp{ps}, respectively. 

Next, a ``measurement update'' step is performed. For legacy \acp{ps}, messages $\gamma_s^{(j)}\big(\underline{\V{x}}_s^{(j)}\!, \underline{r}_s^{(j)}\big)$ passed from $q\big( \underline{\V{x}}^{(j)}_{s}\!, \underline{r}^{(j)}_{s}\!, a^{(j)}_{s}\rmv; \V{z}_{s} \big)$ to $\underline{\V{y}}_s^{(j)}$ are calculated \vspace{0mm}as
\begin{equation}
  \gamma_s^{(j)}\big(\underline{\V{x}}_s^{(j)}\!, 1\big) = \rmv\sum_{a_s^{(j)}\rmv=0}^{M_s} q\big( \underline{\V{x}}^{(j)}_{s}\!, 1, a^{(j)}_{s}\rmv; \V{z}_{s} \big)  \kappa_k^{(j)}\big(a_s^{(j)}\big)
  \label{eq:messageGamma}
  \vspace{0mm}
\end{equation}
and $\gamma_s^{(j)}\big(\underline{\V{x}}_s^{(j)}\!, 0\big) = \gamma_s^{(j)} =  \kappa_s^{(j)}\big(0\big)$.
Finally, beliefs are calculated to approximate the posterior PDFs of \acp{ps}. In particular, for legacy \acp{ps}, beliefs $\tilde{f}\big(\underline{\V{x}}_s^{(j)}\!, \underline{r}_s^{(j)}\big)$ approximating $f\big(\underline{\V{x}}_s^{(j)}\!, \underline{r}_s^{(j)}\rmv\big|\rmv\V{z}_{1:s}\big)$ are obtained\vspace{.5mm} as 
\begin{equation}
  \tilde{f}\big(\underline{\V{x}}_s^{(j)}\!, 1\big) \rmv=\rmv \frac{1}{\underline{C}_s^{(j)}}\alpha_s^{(j)}\big(\underline{\V{x}}_s^{(j)}\!, 1\big)\gamma_s^{(j)}\big(\underline{\V{x}}_s^{(j)}\!, 1\big)
  \label{eq:beliefLegacy}
\end{equation}
and $\tilde{f}\big(\underline{\V{x}}_s^{(j)}\!, 0\big)\rmv=\rmv \underline{f}_s^{(j)} f_{\text{D}}\big(\underline{\V{x}}^{(j)}_{s}\big)$ with\vspace{0mm} $\underline{f}_s^{(j)} \rmv=\rmv \alpha_{\mathrm{n},s}^{(j)}\gamma_s^{(j)} /\underline{C}_s^{(j)}$. The constant $\underline{C}_s^{(j)}$ is given by $\underline{C}_s^{(j)}\rmv\triangleq\rmv \int \alpha_s^{(j)}\big(\underline{\V{x}}_s^{(j)}\!, 1\big)$ $\gamma_s^{(j)}\big(\underline{\V{x}}_s^{(j)}\!, 1\big) \mathrm{d}\underline{\V{x}}_s^{(j)} + \alpha_{\mathrm{n},s}^{(j)}\gamma_s^{(j)}\rmv$. 

\subsection{Particle Approximation of Selected Messages}\label{sec:IV-B}
Note that neither \eqref{eq:messageBeta} \eqref{eq:messageGamma} nor \eqref{eq:singleLikelihood1} \eqref{eq:singleLikelihood2} have a parametric solution in general due to complex integral as well as non-linearity of measurement model \eqref{eq:tau}. As a result, we propose to approximate those messages and likelihood functions by Monte Carlo importance sampling. Let's take \eqref{eq:messageBeta} as an example. If we can approximate the message $\alpha_s^{(j)}\!\big(\underline{\V{x}}_s^{(j)}\!,1\big)$ by a set of weighted particles $\big\{\big(\underline{\V{x}}^{(j,i)}_0,\underline{w}^{(j,i)}_0\big)\big\}_{i=1}^{N_\mathrm{s}}$ in that $\sum_{i=1}^{N_\mathrm{s}}\underline{w}_{0}^{(j,i)} \approx
\int\alpha_s^{(j)}\!\big(\underline{\V{x}}_s^{(j)}\!,1\big)\mathrm{d}\underline{\V{x}}_s^{(j)}$, then we can approximate \eqref{eq:messageBeta} as 
\begin{equation}\label{eq:particleBeta}
  \tilde{\beta}_s^{(j)}\big(a_s^{(j)} = a\big) \rmv=\rmv \rmv\sum_{i=1}^{N_\mathrm{s}} q\big( \underline{\V{x}}^{(j,i)}_0\!, 1, a\rmv; \V{z}_{s} \big)\underline{w}_{0}^{(j,i)} + 1(a) \tilde{\alpha}_{\mathrm{n},s}^{(j)} 
\end{equation}
where $\tilde{\alpha}_{\mathrm{n},s}^{(j)} = 1-\sum_{i=1}^{N_\mathrm{s}}\underline{w}_{0}^{(j,i)}$.

Note that in \eqref{eq:particleBeta} we sample from $\alpha_s^{(j)}\!\big(\underline{\V{x}}_s^{(j)}\!,1\big)$, which can be seen as the prior PDF of legacy \ac{ps} with index $j$. However, note that $q\big( \underline{\V{x}}^{(j)}_{s}\rmv\rmv\rmv,\rmv \underline{r}_s^{(j)}\rmv\rmv\rmv=\rmv1, a^{(j)}_{s}\rmv\rmv; {z}_{s} \rmv\big)$ incorporates the measurement likelihood $f( {z}^{(m)}_s \rmv |\ist \underline{\V{x}}^{(j)}_s)$ (c.f. \eqref{eq:singleLikelihood1}), which suffers from particle degeneracy similarly to \ac{bpf} \cite{BicLiBen:B08}. The underlying inefficiency of sampling from the prior PDF is that few samples fall into the region of high likelihood of $f( {z}^{(m)}_s \rmv |\ist \underline{\V{x}}^{(j)}_s)$ (e.g., the hyperboloid at the first sensor or the intersected curve of two hyperboloids at the second sensor in Fig. \ref{fig:Hyperboloid} \cite{ZhaMey:J24}), which makes a poor approximation of the target message $\beta_s^{(j)}\big(a_s^{(j)}\big)$. Similarly, also the messages $\gamma_{s}^{(j)}\big(\underline{\V{y}}^{(j)}_{s} \big)$ based on \eqref{eq:singleLikelihood1} as well as $\xi^{(m)}_{s}\big(b^{(m)}_{s}\big)$ and $\varsigma^{(m)}_s\big(\overline{\V{y}}^{(m)}_{s} \big)$ based on \eqref{eq:singleLikelihood2} are affected by particle degeneracy). We propose to use \ac{pfl} to overcome particle degeneracy. 

\section{Review of Deterministic and Stochastic \acp{pfl}}
In this section, we review deterministic and stochastic \acp{pfl} and discuss relevant approaches to drive \ac{pfl}-based proposal PDFs\vspace{-2mm}.

\subsection{Deterministic \ac{pfl}: \ac{edh} and \ac{ledh} Flow}
The Bayesian \ac{pfl} \cite{DuaHua:07,DuaHua:09} tries to define a continuous mapping $\Phi: \V{X}_0 \times [0,1] \rightarrow \V{X}_1$ to migrate particles sampled from the topological space of prior PDF $\V{x}^{(i)}_0\sim \V{X}_0$ to the topological space of posterior PDF $\V{x}^{(i)}_1\sim \V{X}_1$. The process is time-dependent w.r.t. a pseudo-time $\lambda\in [0,1]$.  

Let $f(\V{x})$ be the prior PDF, $h(\boldsymbol{x}) \rmv= \rmv f(\boldsymbol{z}|\boldsymbol{x})$ be the likelihood function (e.g. \eqref{eq:tau} for \ac{tdoa} model). Following Bayes' rule, a log-homotopy function is introduced \cite{DuaHua:07,DuaHua:09} \vspace{.5mm} as 
\begin{equation}
  \phi(\boldsymbol{x},\lambda)=\log f(\boldsymbol{x})+\lambda\log h(\boldsymbol{x}).
  \label{homotopyPhi}
  \vspace{.5mm}
\end{equation}
Note that the homotopy function defines a continuous and smooth deformation from $\phi(\boldsymbol{x},0) \rmv=\rmv \log f(\boldsymbol{x})$ to $\phi(\boldsymbol{x},1) \rmv= \log \pi(\boldsymbol{x})$, where $\pi(\boldsymbol{x}) = f(\V{x})f(\V{z}|\V{x})$ is the unnormalized posterior PDF. The drift of the flow 
$\V{\zeta}(\V{x},\lambda)= \text{d}\V{x}/ \text{d}\lambda$
is calculated by solving an \ac{ode} and hence used to drive the dynamics of particles deterministically. 
Since the drift depends on $\V{x}$, the direct integral over $\lambda$ from 0 to 1 typically has no analytical solution. The Euler method is commonly used for numerical implementation, where particle migration is performed by calculating $\V{\zeta}(\boldsymbol{x},\lambda)$ at $N_\lambda$ discrete values of $\lambda$, i.e., $0=\lambda_0<\lambda_1<...<\lambda_{N_\lambda}=1$. First, $N_\mathrm{s}$ particles 
$\big\{\V{x}_{0}^{(i)}\big\}_{i=1}^{N_\mathrm{s}}$
are drawn from $f({\boldsymbol{x}})$. Next, these particles are migrated sequentially across discrete pseudo time steps $l \rmv\in\rmv\{1,\dots,N_{\lambda}\}$,\vspace{1.2mm} i.e.,
\begin{equation}
\boldsymbol{x}^{(i)}_{\lambda_l} = \boldsymbol{x}^{(i)}_{\lambda_{l-1}} + \V{\zeta}(\boldsymbol{x}^{(i)}_{\lambda_{l-1}},\lambda_{l}) (\lambda_{l} - \lambda_{l-1})\hspace{1.5mm}  \label{eq:flow}
\vspace{1mm}
\end{equation}
for all $i\rmv\in\rmv \{1,\dots,N_\mathrm{s}\}$. In this way, particles 
$\{\V{x}_1^{(i)}\}_{i=1}^{N_\mathrm{s}}$
representing the posterior PDF $\pi(\boldsymbol{x})$ are obtained.

If $\log f(\boldsymbol{x})$ and $\log h(\boldsymbol{x})$ are polynomials (e.g., $f(\boldsymbol{x})$ and $h(\boldsymbol{x})$ are Gaussians or in another exponential family), the drift term can be solved exactly \cite{DuaHua:10}. In particular, let us consider a Gaussian prior $f(\V{x}) = \Set{N}(\V{x}; \V{\mu}_0,\M{P})$ with mean ${\boldsymbol{\mu}}_0$ and covariance matrix $\M{P}$ as well as a linear measurement model $\V{z} = \M{H} \V{x} + \V{v}$. Here, the measurement noise $\V{v}$ is zero-mean Gaussian with covariance matrix $\M{R}$. The \ac{edh} flow \cite{DuaHua:10} is given \vspace{1.2mm} by 
\begin{equation}
  \V{\zeta}_{\text{d}}(\boldsymbol{x},\lambda) = \M{A}(\lambda)\boldsymbol{x} + \V{b}(\lambda) 
  \vspace{0mm}
\label{eq:affineTransform}
\end{equation}
where
\vspace{1mm}
\begin{align}
  \M{A}(\lambda) &= -\frac{1}{2}\M{P}\M{H}^\T(\lambda \M{H}\M{P}\M{H}^\T+\M{R})^{-1}\M{H} \nn\\[1mm]
  \V{b}(\lambda) &= (\M{I}+2\lambda \M{A}(\lambda))\big[(\M{I}+\lambda \M{A}(\lambda))\M{P}\M{H}^\T \M{R}^{-1}\boldsymbol{z}+\M{A}(\lambda){\boldsymbol{\mu}}_0\big].  \nn\\[-2.5mm] 
  \nn
\end{align}
For nonlinear measurement models $\V{z} = \M{h} (\V{x}) + \V{v}$, a suboptimal linearization step is employed, i.e. $\tilde{\M{H}}_{{l-1}} = \frac{\partial\M{h}}{\partial\V{x}}\big|_{\V{x}=\V{\mu}_{\lambda_{l-1}}}$  \vspace{-.3mm} at a current mean $\V{\mu}_{\lambda_{l-1}}$. This mean is propagated in parallel to the particles $\{\V{x}_{\lambda_{l-1}}^{(i)}\}_{i=1}^{N_\mathrm{s}}$ by using \eqref{eq:flow}. Alternatively, linearization can be performed\vspace{-1mm} at each particle, i.e., at $\tilde{\M{H}}^{(i)}_{{l-1}} = \frac{\partial\M{h}}{\partial\V{x}}\big|_{\V{x}=\V{x}^{(i)}_{\lambda_{l-1}}}$ where $\tilde{\M{A}}^{(i)}(\lambda)$ and $\tilde{\V{b}}^{(i)}(\lambda)$ are computed for each particle respectively. This is known as the \ac{ledh} flow. The \ac{edh} flow is computationally much faster than the \ac{ledh} flow since it only calculates a single pair of global flow parameters $\M{A}$ and $\V{b}$. However, the PDF of particles after the flow still follows a Gaussian PDF. (Note that the flow in \eqref{eq:affineTransform} can be seen as a sequence of affine transformations applied to the Gaussian prior.) On the other hand, the \ac{ledh} flow provides particles that follow an arbitrary probability PDF since each particle follows an individual affine transform. However, since a different pair of flow parameters is computed for each particle, the computational complexity of the \ac{ledh} flow is $N_\mathrm{s}$ times higher than \ac{edh} flow.

\subsection{Stochastic \ac{pfl} Gromov's Flow}\label{subsec:Gromov}
The migration of particles corresponding to \eqref{homotopyPhi} can alternatively be describe by a \ac{sde} \cite{DuaHuaNou:18}, i.e., 
\begin{equation}
    \mathrm{d} \V{x} = \V{\zeta}_{\mathrm{s}}(\V{x}, \lambda) \mathrm{d}\lambda + \mathrm{d} \V{w}_\lambda
    \label{eq:stochasticFlow}
\end{equation}
where $\V{\zeta}_{\mathrm{s}}(\V{x}, \lambda) \in \mathbb{R}^N$ is the stochastic drift and $\mathrm{d} \V{w}$ is the $N$-dimensional standard Brownian motion with positive-definite diffusion matrix $\mathrm{d} \V{w} \ist \mathrm{d} \V{w}^{\T} = \M{Q}(\lambda) \in \mathbb{R}^{N\times N}$~\cite{DuaHuaNou:18}.
The stochastic \ac{pfl} outperforms deterministic \ac{pfl} for improving the transient dynamics. In some applications, their random nature can avoid numerical issues, referred to as stiffness of the flow \cite{DaiDau:22}), related to implementing the flow using discrete values of pseudo-time $\lambda$.

One of the most popular stochastic flow with a closed-form solution is Gromov's flow, where
\begin{equation}\label{deterministicDrift}
  \V{\zeta}_{\mathrm{g}}(\V{x}, \lambda) = -\big(\nabla_{\rmv\V{x}}\nabla_{\rmv\V{x}}^\T \phi\big)^{-1} \nabla_{\rmv\V{x}} \log h
\end{equation}
with the following solution for diffusion matrix
\begin{equation}\label{diffusionMatrix}
\M{Q}_\text{g} = -\big(\nabla_{\rmv\V{x}}\nabla_{\rmv\V{x}}^\T \phi\big)^{-1} \big(\nabla_{\rmv\V{x}}\nabla_{\rmv\V{x}}^\T \log h\big)\big(\nabla_{\rmv\V{x}}\nabla_{\rmv\V{x}}^\T \phi\big)^{-1}.
\end{equation}
One of the advantages of Gromov's flow is that it provides an unbiased estimate of the state~\cite{Cro-19}.

Given Gaussian prior and linear approximation of measurement model as in the \ac{edh} flow, we have $\nabla_{\rmv\V{x}} \log h = \M{H}^\T \M{R}^{-1}(\boldsymbol{z}-\M{H}\boldsymbol{x})$, $\nabla_{\rmv\V{x}}\nabla_{\rmv\V{x}}^\T \log h = -\M{H}^\T \M{R}^{-1}\M{H}$ and $\nabla_{\rmv\V{x}}\nabla_{\rmv\V{x}}^\T \phi = -\M{P}^{-1} - \lambda \M{H}^\T \M{R}^{-1}\M{H}$. Then we can rewrite \eqref{deterministicDrift} and \eqref{diffusionMatrix} as 
\begin{equation}\label{driftGaussian}
  \V{\zeta}_{\mathrm{g}}(\V{x}, \lambda) = \Big(\M{P}^{-1} + \lambda \M{H}^\T \M{R}^{-1}\M{H}\Big)^{-1} \M{H}^\T \M{R}^{-1}(\boldsymbol{z}-\M{H}\boldsymbol{x})
\end{equation}
\begin{equation}\label{diffusionGaussian}
     \M{Q}_\text{g} = (\M{P}^{-1} \rmv\rmv\rmv+\rmv\rmv \lambda \M{H}^\T \M{R}^{-1}\M{H})^{-1}(\M{H}^\T \M{R}^{-1}\M{H})(\M{P}^{-1} \rmv\rmv\rmv+\rmv\rmv \lambda \M{H}^\T \M{R}^{-1}\M{H})^{-1}.
\end{equation}
We can still parameterize the drift similarly as \ac{edh} as 
\begin{equation}
  \V{\zeta}_{\text{g}}(\boldsymbol{x},\lambda) = \M{A}_{\text{g}}(\lambda)\boldsymbol{x} + \V{b}_{\text{g}}(\lambda)  \nn
  \vspace{0mm}
\end{equation}
where the parameters of the flow are given by
\vspace{1mm}
\begin{align}
  \M{A}_{\text{g}}(\lambda) &= -\Big(\M{P}^{-1} + \lambda \M{H}^\T \M{R}^{-1}\M{H}\Big)^{-1} \M{H}^\T \M{R}^{-1}\M{H} \nn\\[1mm]
  \V{b}_{\text{g}}(\lambda) &= \Big(\M{P}^{-1} + \lambda \M{H}^\T \M{R}^{-1}\M{H}\Big)^{-1} \M{H}^\T \M{R}^{-1}\boldsymbol{z}.  \nn\\[-2.5mm] 
  \nn
\end{align}
\subsection{\ac{pfl} as Proposal PDF}\label{subsec:invertibleMapping}
Although \ac{pfl} methods display very promising characteristics, due to approximations related to the discretization of pseudo time $\lambda$, they are unable to provide an asymptotically optimal representation of the posterior PDF $\pi(\V{x})$. However, for asymptotically optimal estimation, \ac{pfl} methods can be used to provide samples and a proposal PDF $q(\V{x})$ for importance sampling \cite{AruMasGorCla:02}. 

Direct evaluation of the proposal PDF $q(\V{x})$ related to \ac{pfl} is complicated by the fact that a closed-form expression of $q(\V{x})$ is unavailable. For the numerical evaluation of $q({\V{x}}^{(i)}_{1})$, $i = 1,\dots,N_\mathrm{s}$, the following two approaches can be employed:
i) deriving an invertible mapping from the prior to the posterior PDF \cite{LiCoates:17} and ii) developing a Gaussian proposal PDF where mean and covariance are updated sequentially based on flow equations \cite{LiZhaoCoates:15}. 

As presented in \cite{LiCoates:17}, for an affine deterministic flow, there exists an invertible mapping $\theta: f({\V{x}}^{(i)}_{0})\rightarrow \pi({\V{x}}^{(i)}_{1})$ with linear operator $\theta$. This motivates a proposal PDF given by  $q({\V{x}}^{(i)}_{1}) = f({\V{x}}^{(i)}_{0}) / \tilde{\theta}$ where the approximate mapping factor $\tilde{\theta}$ is defined\vspace{1mm} as
\begin{equation}
\tilde{\theta} = \prod_{l=1}^{N_\lambda} \big| \ist \mathrm{det}\big[\V{I}+(\lambda_{l} - \lambda_{l-1}) \ist \tilde{\M{A}}(\lambda_l)\big] \big|. 
\label{eq:mappingFactor}
\vspace{1mm}
\end{equation} 
Deriving an invertible mapping for stochastic flows remains an open research problem. 
As an alternative, following the second approach discussed above, we develop a Gaussian proposal PDF based on stochastic flow, as discussed next.

\section{BP-Based Source Localization With Embedded Stochastic \ac{pfl}}

In this section, we develop a  \ac{gmm} representation of posterior PDFs based on stochastic \ac{pfl} and demonstrate how it can be incorporated into our BP framework for source localization\vspace{-1mm}.

\subsection{Proposal Evaluation and \ac{gmm}-Based Representation}\label{subsec:GMM}
Proposal evaluation based on deterministic flow within a \ac{gmm} representation of beliefs has been presented in~\cite{ZhaMey:J24}. Here, we adapt this strategy to the use of a stochastic flow within the \ac{gmm} representation. The primary motivation for using a \ac{gmm} representation is that it can represent non-Gaussian beliefs with potentially complicated shapes. Being able to model complicated beliefs is important in multisensor localization and tracking problems where the dimension of the individual sensor measurements is lower than the dimension of object positions as, e.g., in the scenario shown in Fig.~\ref{fig:Hyperboloid}. Moreover, by leveraging a mixture of Gaussian kernels to model beliefs, \ac{pfl} can be performed for each Gaussian kernel in parallel.

Let's first consider a single Gaussian prior $f(\V{x}) = \Set{N}(\V{x}; \V{\mu}_0, \M{\Sigma}_0 \triangleq \M{P})$. For discrete time $0 = \lambda_0 < \dots < \lambda_{N_\lambda} = 1$, we can update the mean and covariance by making using of the stochastic flow in~\eqref{eq:stochasticFlow}, i.e.,
\begin{equation}
\V{\mu}_{\lambda_l} = \V{\mu}_{\lambda_{l-1}} + \V{\zeta}_\mathrm{g}(\V{\mu}_{\lambda_{l-1}},\lambda_l)(\lambda_l - \lambda_{l-1})
\label{meanTransform}
\end{equation}
\begin{align}
\hspace{0mm}\M{\Sigma}_{\lambda_l} = \ist\ist & [\M{I}+(\lambda_{l}-\lambda_{l-1})\M{A}_{\text{g}}(\lambda_l)]\V{\Sigma}_{\lambda_{l-1}}[\M{I}+(\lambda_{l}-\lambda_{l-1})\M{A}_{\text{g}}(\lambda_l)]^{\text{T}} \nn\\[1mm]
                                      & + (\lambda_l - \lambda_{l-1})\M{Q}_{\text{g}}(\lambda_l) \label{covTransform}    \\[-5mm]
\nn                   
\end{align}
for $l = 1,\dots,N_\lambda$. We can then evaluate the proposal at the particles as $q(\V{x}_1^{(i)}) = \Set{N}(\V{x}_1^{(i)}; \V{\mu}_1, \M{\Sigma}_1)$. 

Next, we utilize \eqref{meanTransform} and \eqref{covTransform} in a \ac{gmm} model. More precisely, instead of sampling $\big\{\V{x}_{0}^{(i)}\big\}_{i=1}^{N_\mathrm{s}}$ from a Gaussian prior $f(\V{x}) = \Set{N}(\V{x}; \V{\mu}_0,\M{\Sigma}_0)$, we model the prior as \ac{gmm} with $N_{\mathrm{k}}$ kernels, i.e., $f(\V{x}) \propto \sum_{k=1}^{N_\mathrm{k}}\Set{N}(\V{x}; \V{\mu}_0^{(k)},\M{\Sigma}_0^{(k)})$. This mixture representation makes it possible to approximate arbitrary PDFs. In particular, it enables to closely approximate localization information on a hyperboloid as provided by \ac{tdoa} measurements. Then for each kernel $k$, we sample $\big\{\V{x}_{0}^{(i,k)}\big\}_{i=1}^{N_\mathrm{p}}$ from $\Set{N}(\V{x}; \V{\mu}_0^{(k)},\M{\Sigma}_0^{(k)})$ and use $\big\{\big\{\V{x}_{0}^{(i,k)}\big\}_{i=1}^{N_\mathrm{p}}\big\}_{k=1}^{N_\mathrm{k}}$ as the particle set. Hence, the equivalent particle number is $N_\mathrm{p}\times N_\mathrm{k}$. 
  For the stochastic flow and using the Gaussian solution, the proposal at particles for each kernel is 
 $q(\V{x}_1^{(i,k)}) = \Set{N}(\V{x}_1^{(i,k)}; \V{\mu}_1^{(k)}, \M{\Sigma}_1^{(k)})$.

The methodology for updating kernels from particles within the \ac{gmm} framework is discussed in~\cite{ZhaMey:J24}. In particular, one flow is performed for each Gaussian kernel in the  \ac{gmm}, and the resulting particles are used to compute updated Gaussian parameters. The number of Gaussian kernels in our proposed \ac{gmm} algorithm is much smaller than the number of particles, i.e., $N_\mathrm{k}<<N_\mathrm{p}$. As a result, compared to \cite{LiZhaoCoates:15, PalCoa:19}, much fewer flow parameters have to be computed, which results in a significantly reduced computational complexity.

\subsection{BP With Embedded Stochastic \ac{pfl}}
We finally discuss how to approximate BP messages in Section \ref{sec:IV-B} based on stochastic \ac{pfl}. Let's still use \eqref{eq:messageBeta} as an example. In addition, for simplicity, let's assume that the message $\alpha_s^{(j)}\!\big(\underline{\V{x}}_s^{(j)}\!,1\big)$ is represented by a single Gaussian $ \Set{N}( \underline{\V{x}}_s^{(j)} ; \V{\mu}_s^{(j)}, \M{\Sigma}_s^{(j)})$ and approximate existent probability $\tilde{\alpha}_{\mathrm{e},s}^{(j)} \approx \int \alpha_s^{(j)}\!\big(\underline{\V{x}}_s^{(j)}\!,1\big) \mathrm{d} \underline{\V{x}}_s^{(j)}$. We can then directly sample from the this Gaussian to obtain a set of weighted particles $\big\{\big(\underline{\V{x}}^{(j,i)}_{s}\rmv\rmv,\underline{w}^{(j,i)}_{s}\big)\big\}_{i=1}^{N_\mathrm{s}}$ with $\underline{w}^{(j,i)}_{s} = \tilde{\alpha}_{\mathrm{e},s}^{(j)}/ N_{\mathrm{s}}$ as an alternative representation of $\alpha_s^{(j)}\!\big(\underline{\V{x}}_s^{(j)}\!,1\big)$. To improve the sampling efficiency related to the computation of the messages ${\beta}_s^{(j)}\big(a_s^{(j)} = m\big), m \in \{1,\dots,M_s\}$, we migrate the samples $\big\{\underline{\V{x}}^{(j,i)}_{s}\big\}_{i=1}^{N_\mathrm{s}}$ using stochastic \ac{pfl}. In particular, we perform $M_s$ parallel flows, one for each measurement $z^{(m)}_s\rmv\rmv\rmv$, $m \rmv\in\rmv \{1,\dots,M_s\}$. After the $M_s$ parallel flows have been completed, we computed $M_s$ sets of particles, i.e., $\big\{\underline{\V{x}}^{(j,i)}_{s,m}\big\}_{i=1}^{N_\mathrm{s}}$, $m \rmv\in\rmv \{1,\dots,M_s\}$. Then for $a = 1,\dots,M_s$, we can rewrite \eqref{eq:particleBeta} as 
\begin{equation}\label{eq:particleFlowBeta}
  \tilde{\beta}_s^{(j)}\big(a_s^{(j)} = a\big) \rmv=\rmv \rmv\sum_{i=1}^{N_\mathrm{s}} q\big( \underline{\V{x}}^{(j,i)}_{s,a}\!, 1, a; \V{z}_{s} \big)\underline{w}_{s,a}^{(j,i)} + 1(a) \ist \tilde{\alpha}_{\mathrm{n},s}^{(j)}
\end{equation}
with $\tilde{\alpha}_{\mathrm{e},s}^{(j)} = 1 - \tilde{\alpha}_{\mathrm{n},s}^{(j)} $ (cf.~\eqref{eq:particleBeta}). Here, particle weights are computed as  $\underline{w}_{s,a}^{(j,i)} = \Set{N}\big( \underline{\V{x}}_s^{(j)} ; \V{\mu}_s^{(j)}, \M{\Sigma}_s^{(j)} \big) /$ $\Set{N}\big( \underline{\V{x}}_s^{(j)} ; \V{\mu}_{s,a}^{(j)}, \M{\Sigma}_{s,a}^{(j)}\big) \ist \underline{w}_{s}^{(j,i)}\rmv\rmv$ where the parameters of the Gaussian proposal,  i.e., $\V{\mu}_{s,a}^{(j)}$ and $\M{\Sigma}_{s,a}^{(j)}$ are obtained following \eqref{meanTransform} and \eqref{covTransform}.

If the message $\alpha_s^{(j)}\!\big(\underline{\V{x}}_s^{(j)}\!,1\big)$ is represented by a \ac{gmm}, \eqref{eq:particleFlowBeta} can be rewrite as 
\begin{equation}\label{eq:particleFlowBetaGMM}
  \tilde{\beta}_s^{(j)}\big(a_s^{(j)} = a\big) \rmv\rmv=\rmv\rmv \rmv\frac{1}{N_\mathrm{k}}\sum_{k=1}^{N_\mathrm{k}} \rmv \sum_{i=1}^{N_\mathrm{p}} q\big( \underline{\V{x}}^{(j,i,k)}_{s,a}\!, 1, a; \V{z}_{s} \big)\underline{w}_{s,a}^{(j,i,k)} \rmv + \rmv 1(a) \ist \tilde{\alpha}_{\mathrm{n},s}^{(j)} \nn
\end{equation}
where $\underline{w}_{s,a}^{(j,i,k)}$ is computed based on the Gaussian proposal resulting from the flow performed for measurement $m\rmv=\rmv a$ and Gaussian kernel $k$. Following a similar flow-based processing, further messages for legacy and new \acp{ps}, namely $\gamma_{s}^{(j)}\big(\underline{\V{y}}^{(j)}_{s} \big)$, $\xi^{(m)}_{s}\big(b^{(m)}_{s}\big)$ and $\varsigma^{(m)}_s\big(\overline{\V{y}}^{(m)}_{s} \big)$ can also be efficiently computed. As discussed in the next section, by embedding \ac{pfl}, BP message passing can provide accurate estimation results in challenging nonlinear and high-dimensional problems even with a relatively moderate number of particles.

\section{Implementation Aspects and Results}
\label{sec:simRes}

In this section, we test our proposed BP algorithm with \ac{gmm} \ac{pfl} embedded for sequential processing of 3-D \ac{tdoa} measurements in a 3-D multisource localization problem. To the best of our knowledge, the \ac{gmm} \ac{pfl} algorithm hasn’t been tested for a challenging task to detect and localize multiple sources appearing at the same time in a 3-D space. As expected from the discussion in Section \ref{subsec:Gromov} and \ref{subsec:GMM}, we will demonstrate that the \ac{gmm} Gromov's flow outperforms the other reference methods.
The baseline method under comparison is the bootstrap particle filter, wherein the prior message is utilized for particle sampling (abbreviated as ``PM" for later notation). Apart from the Gromov's flow, we also implement \ac{gmm} with deterministic flows embedded as a reference, such as the \ac{edh} flow \cite{DuaHua:10} (abbreviated as ``EDH") and the \ac{ledh} flow (abbreviated as ``LEDH").

\subsection{Simulation Parameters}
\begin{figure*}[ht!]
  \centering
  \hspace*{1mm}\begin{minipage}[H!]{0.23\textwidth}
  \centering
  \subcaptionbox{}{\hspace{-6mm}\includegraphics[width=1.15\textwidth]{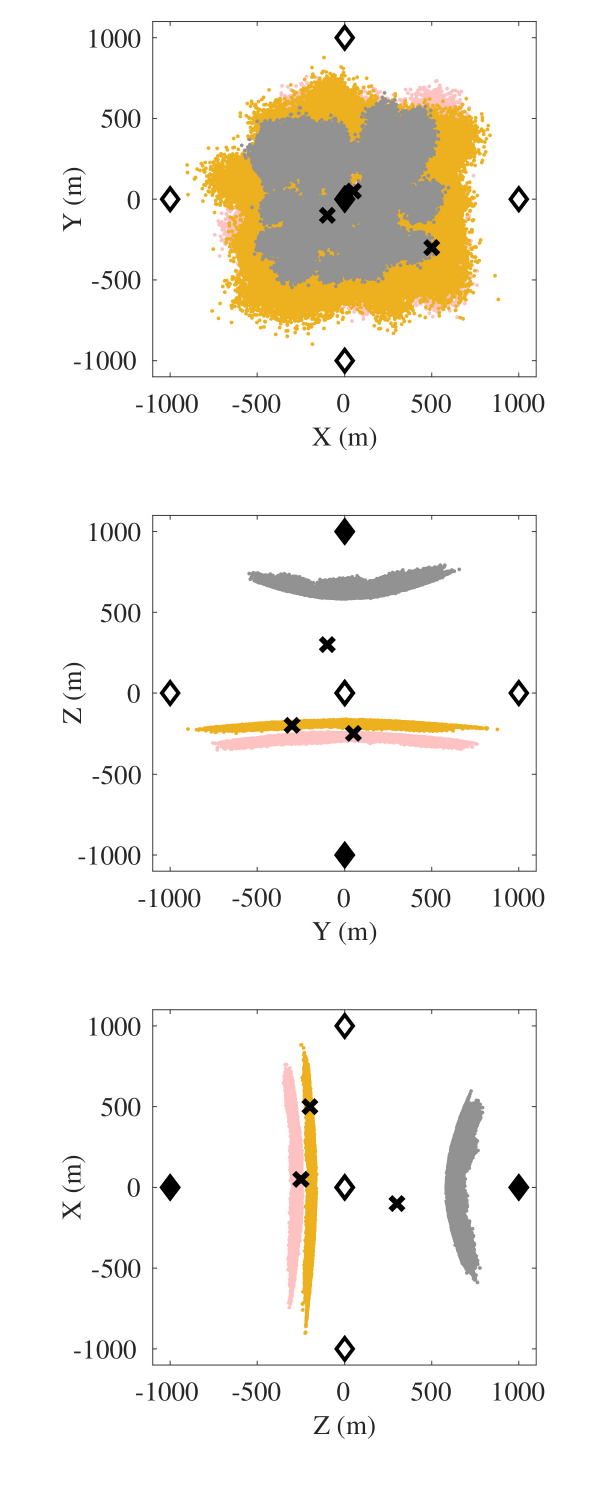}\vspace{-3mm}}
  \end{minipage}\hspace{1mm}
  \begin{minipage}[H!]{0.23\textwidth}
  \centering
  \subcaptionbox{}{\hspace{-6mm}\includegraphics[width=1.15\textwidth]{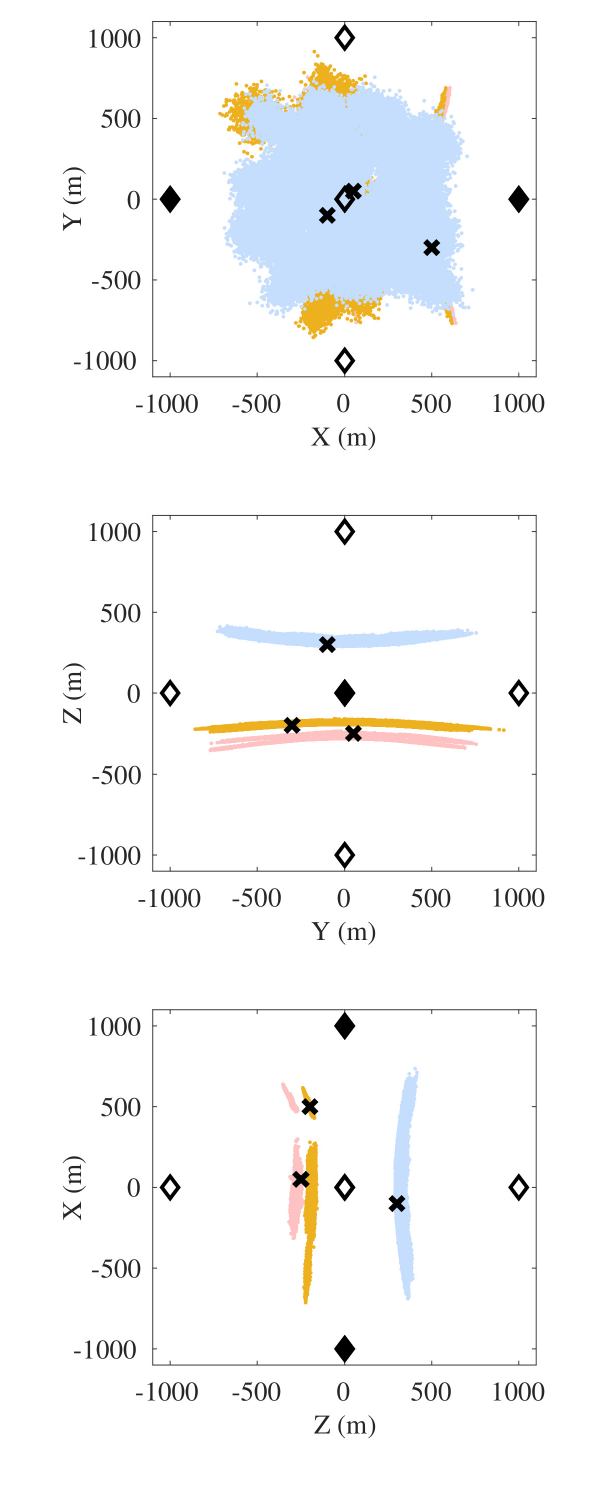}\vspace{-3mm}}
  \end{minipage}\hspace{1mm}
  \begin{minipage}[H!]{0.23\textwidth}
  \centering
  \subcaptionbox{}{\hspace{-6mm}\includegraphics[width=1.15\textwidth]{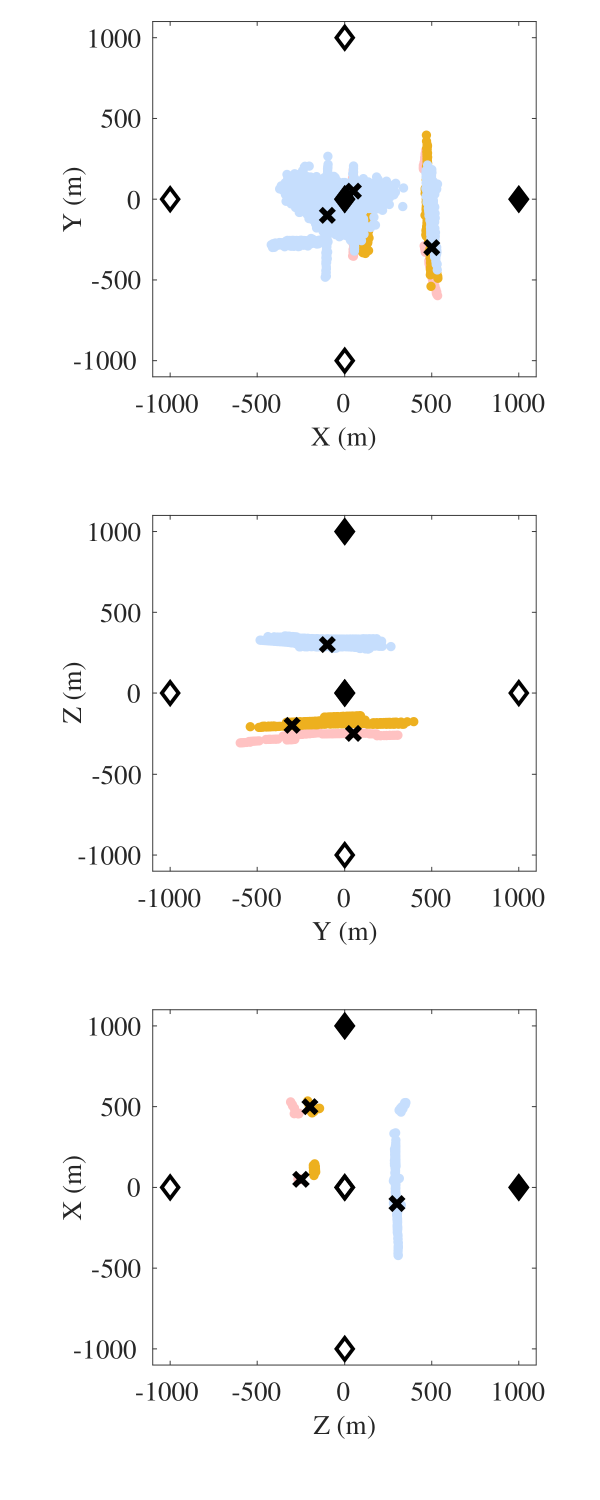}\vspace{-3mm}}
  \end{minipage}\hspace{1mm}
  \begin{minipage}[H!]{0.23\textwidth}
  \centering
  \subcaptionbox{}{\hspace{-6mm}\includegraphics[width=1.15\textwidth]{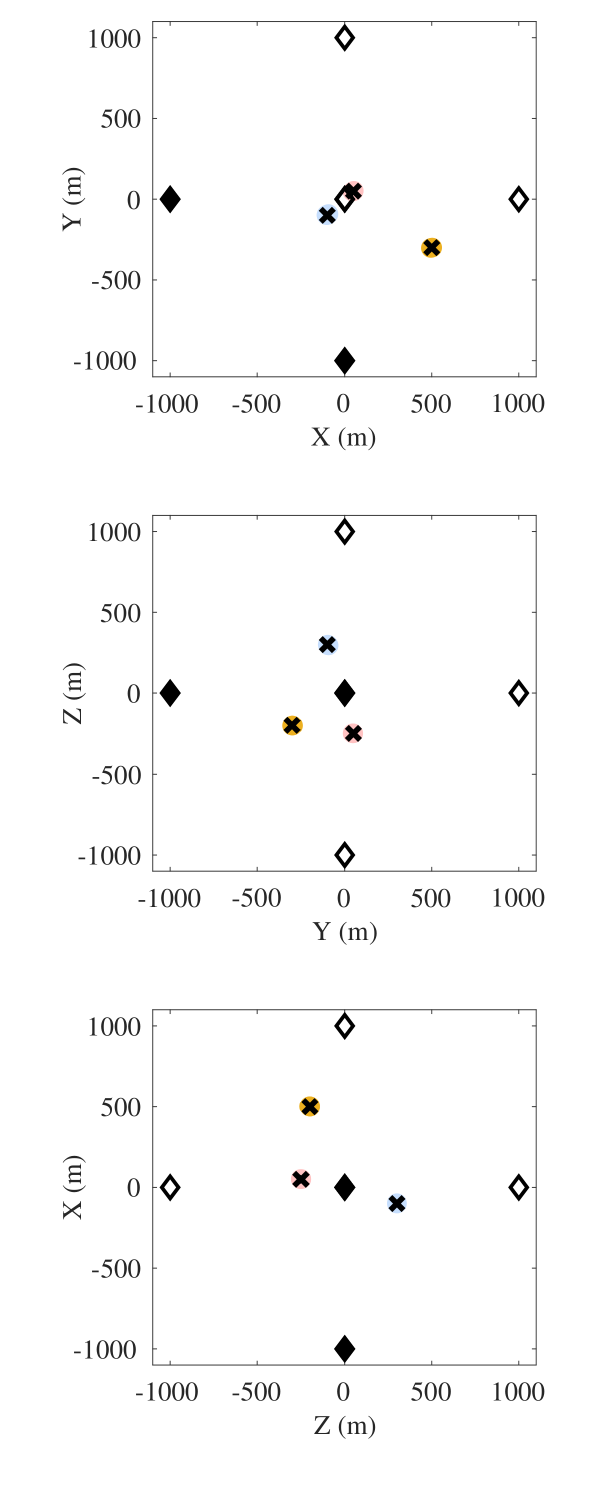}\vspace{-3mm}}
  \end{minipage}
  \caption{Particle-based representation after the \ac{edh} flow of the spatial PDFs by \ac{gmm} model. There exist three signal sources in a 3-D environment with clutters, and the state is sequentially filtered by sensor: (a) $s=1$, (b) $s=2$, (c) $s=4$, and (d) $s=9=n_{\text{s}}$. Filled diamonds indicate the locations of receivers involved in the current update step; the locations of other receivers are indicated by empty diamonds. X-marks indicate the sources' ground truth positions.
  }
  \label{fig:simres}
  \vspace{0mm}
\end{figure*} 

In our simulations the sources are randomly placed on a \ac{roi} of $[-1000\ist\text{m}, \ist 1000\ist\text{m} ]  \times [-1000\ist\text{m}, \ist 1000\ist\text{m}] \times [-1000\ist\text{m}, \ist 1000\ist\text{m}]$. 
The source-originated TDOAs $\rv{z}^{(m)}_{kl}$ at receiver pair $(k,l)$ are distributed according to \eqref{eq:tau} with a standard deviation of $\sigma_z = 0.001$m$/c$ for the noise $v^{(m)}_{kl}$, where $c$ is the propagation speed of signal in the environment. We set $c=1500$m/s to model acoustic propagation in water. The clutter PDF $f_{\text{c}}\big( {z}^{(m)}_{s} \rmv\big)$ at receiver pair $(k,l)$ is uniform on $\|{\V{q}}^{(k)} - {\V{q}}^{(l)}\|/c$ following Poisson point process with mean $\mu_{\text{c}}=1$.
Six receivers are located at the center of each face of the \ac{roi} cube. The probability of detection $p_{\text{d}}$ is set to $0.95$. The number of receiver pairs $S$ is $9$. 

For a single realization, Fig. \ref{fig:simres} shows simulation results for a scenario with three static sources. For each source, the \ac{gmm} consists of $N_\mathrm{k}=100$ kernels; each is importance sampled using \ac{edh} flow. The particle size is $\underline{N}_\mathrm{p} = 1000$ for previously detected sources and $\overline{N}_\mathrm{p} = 1000$ for newly detected sources at the current sensor. 
In Fig. \ref{fig:simres}, particles are shown in different colors to represent the spatial PDFs of the \acp{ps} $j$ with the existence probabilities $f(r_s^{(j)} \rmv=\rmv 1| \V{z}_{1:s})$ larger than $P_{\text{th}}=0.5$ (hence regarded as existence) by sensor $s$, i.e., $f\big(\V{x}_s^{(j)} \big| r_s^{(j)} \rmv=\rmv 1, \V{z}_{1:s}\big)$ for $s = 1, 2, 4, 9$ in Fig. \ref{fig:simres}(a)-(d) respectively.  Some observations regarding the evolution of particles between sensor pairs are as\vspace{.5mm} follows.
\begin{itemize}
\item The first receiver pair ($s=1$) has one missed detection and one clutter measurement. Two sets of particles (pink and orange) meet the true locations of two sources, while the other set (grey) corresponds to the clutter. 
Spatial PDFs are approximated hyperbolas in Y-Z and X-Z planes. The particles in the X-Y plane have not been updated effectively since the first sensor pair is at unfavorable locations for localization in that plane\vspace{1mm}.

\item The second receiver pair ($s=2$) has no missed detection. The two components with the highest existence probabilities are the legacy components generated from the measurements of the first receiver pair (pink and orange). 
The third existing component (blue) is regarded as the third source, which is newly detected. Note that the clutter source shown as a grey particle in the first receiver pair no longer exists since there is no measurement matching that clutter in the current receiver pair\vspace{1mm}.

\item The number of sources has been correctly determined after the first two sensors. After the fourth sensor had been updated, spatial PDFs of the three sources were still multimodal but more concentrated around the true source locations than the $s=2$ result. Their spatial PDFs have modes that correspond to the intersection points of the hyperbolas related to previous measurements and current measurements\vspace{1mm}.
\item After the update step has been performed for all receiver pairs, all three existing components have a single mode well localized around the actual locations of the three sources. 
\end{itemize}

\subsection{Results}
To investigate how the performance of our method depends on \ac{gmm} parameters and sampling size, we simulated scenarios with different numbers of kernels and particles per kernel. 

Table I shows the mean optimum subpattern assignment (OSPA) error \cite{SchVoVo:J08} (with a cutoff threshold at 50) and runtime per run for different algorithms and system parameters. Note that for PM, we have $N_{\mathrm{p}} \rmv=\rmv N_{\mathrm{s}}$. All methods are simulated on a single core of an Intel Xeon Gold 5222 CPU. The runtime is averaged over 100 Monte Carlo runs, each with five static sources randomly placed in \ac{roi}. It can be seen that the PM performed the worst in terms of OSPA error. This is expected since the PM is based on bootstrap sampling, which suffers particle degeneracy in the considered scenario.
In contrast, the implementation based on deterministic flow, such as EDH, can achieve a much smaller OSPA error with much fewer particles and reasonable runtime. The implementation based on the LEDH has a slightly better performance than EDH, but the runtime has been increased significantly (see ID 3 vs. ID 4). Further increasing particle size $\underline{N}_\mathrm{p}$ and $\overline{N}_\mathrm{p}$ can improve the localization accuracy (see ID 5 and 6 vs. ID 3) but limited to a bottleneck with OSPA around 20. The implementation based on stochastic flow such as Gromov's flow can significantly reduce the OSPA error compared with EDH flow using same particle size (see ID 7 vs. ID 3, and ID 8 vs. ID 5), whereas three times in runtime. Overall, Gromov's flow with parameter setting as ID 7 is best among all tested methods with a trade-off between localization accuracy and runtime.

\begin{table}[ht!]
  \centering 
  \renewcommand{\arraystretch}{1.25}
  \begin{tabular}{ C{0.3cm} | C{1.6cm} | C{2cm} | C{1.2cm} | C{1.4cm} }
  \hline
  ID & Method & $(N_\mathrm{k},\underline{N}_\mathrm{p},\overline{N}_\mathrm{p})$ & OSPA & Runtime(s)\\[.3mm]
  \hline
  1 & PM & $(-,2\text{e}6,2\text{e}6)$ & $43.90$ & $75.4$ \\
  2 & PM & $(-,1\text{e}7,1\text{e}7)$ & $32.07$ & $443.3$\\[-.3mm]
  3 & EDH & $(100,500,30)$ & $25.17$ & $196.9$ \\[-.3mm]
  4 & LEDH & $(100,500,30)$ & $23.23$ & $4934.2$ \\[-.3mm]
  5 & EDH & $(100,3\text{e}3,500)$ & $20.57$ & $379.6$ \\[-.3mm]
  6 & EDH & $(100,1\text{e}4,1\text{e}4)$ & $19.58$ & $2586.8$ \\[-.3mm]
  7 & Gromov & $(100,500,30)$ & $10.43$ & $568.8$ \\[-.3mm]
  8 & Gromov & $(100,3\text{e}3,500)$ & $8.75$ & $1356.1$ \\[-.3mm]
  \hline
  \end{tabular}
  \vspace{0mm}
  \caption{Simulated mean OSPA error and runtime per run for different algorithms and system parameters.}
  \vspace{-5mm}
\end{table}
\vspace{3mm}

\begin{figure}[ht!]
  \centering
  \hspace{-6mm}\includegraphics[scale=1]{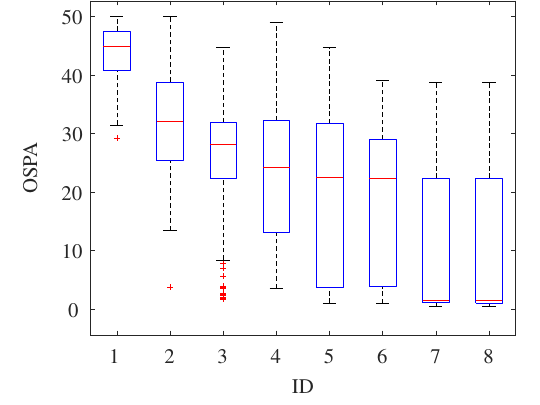}
  \caption{Statistics of the OSPA error for different algorithms. Each column corresponds to the algorithm with the same ID as in Table I.}
  \label{fig:MOSPAresults}
  \vspace{-1mm}
\end{figure}

The statistical box plot \cite{Tuk:B77} of the 100 OSPA samples for each algorithm is shown in Fig. \ref{fig:MOSPAresults} for further comparison. Here, the box with a blue boundary has its lower and upper bound as the first and third quartile of the 100 samples, respectively. The vertical range of the box is the \ac{iqr}. The red line in the box is the median. The black lines that extend from the box are the expected variation as either the minimum/maximum value of the samples or the whiskers (i.e., $1.5\times$ \ac{iqr} extension from the box boundary) if there are outliers. If the second case, the outliers are marked as red `$+$'s. Notably, the median of Gromov's flow is much lower than that of EDH flow, which means it has a much higher chance of detecting all five static\vspace{-2mm} sources.

\section{Conclusion}
In this paper, we propose a BP method for localizing an unknown number of sources in 3-D based on TDOA measurements. Our method relies on \acp{gmm} to represent approximate marginal posterior PDFs, called beliefs, with complicated shapes for accurate localization in 3-D using a nonlinear measurement model.  By passing BP messages, represented as weighted particles, over a factor graph, our method can simultaneously detect and localize an unknown number of sources in the presence of \ac{da} uncertainty, false alarms, and clutter.  The novelty of our approach lies in leveraging stochastic \ac{pfl} for the efficient update of GMM parameters of beliefs. The considered stochastic \ac{pfl}, referred to as Gromov's flow, achieves the best accuracy-complexity tradeoff in numerical experiments, compared to previously considered deterministic flows. Future research includes information fusion of multiple types of measurement models with \ac{pfl} and the development of novel stochastic \acp{pfl} with improved transient dynamics. Modeling unresolved measurements \cite{KocVan:J97}, combining \ac{pfl} with BP-based track-before-detect methods \cite{LiaKroMey:J23}, or introducing deep learning techniques to refine the BP solution \cite{LiaMey:J24} are also possible venues for future \vspace{-2mm} research.

\section*{Acknowledgment}
The material presented in this work was supported by Qualcomm Innovation Fellowship 492866 and by the Office of Naval Research under Grant N00014-23-1-2284.

\renewcommand{\baselinestretch}{1.03}
\bibliographystyle{IEEEtran}
\bibliography{StringDefinitions,IEEEabrv,references,Books}

\begin{thebibliography}{10}
\providecommand{\url}[1]{#1}
\csname url@samestyle\endcsname
\providecommand{\newblock}{\relax}
\providecommand{\bibinfo}[2]{#2}
\providecommand{\BIBentrySTDinterwordspacing}{\spaceskip=0pt\relax}
\providecommand{\BIBentryALTinterwordstretchfactor}{4}
\providecommand{\BIBentryALTinterwordspacing}{\spaceskip=\fontdimen2\font plus
\BIBentryALTinterwordstretchfactor\fontdimen3\font minus
  \fontdimen4\font\relax}
\providecommand{\BIBforeignlanguage}[2]{{%
\expandafter\ifx\csname l@#1\endcsname\relax
\typeout{** WARNING: IEEEtran.bst: No hyphenation pattern has been}%
\typeout{** loaded for the language `#1'. Using the pattern for}%
\typeout{** the default language instead.}%
\else
\language=\csname l@#1\endcsname
\fi
#2}}
\providecommand{\BIBdecl}{\relax}
\BIBdecl

\bibitem{JanMeySny:J23}
J.~Jang, F.~Meyer, E.~R. Snyder, S.~M. Wiggins, S.~Baumann-Pickering, and J.~A.
  Hildebrand, ``Bayesian detection and tracking of odontocetes in 3-d from
  their echolocation clicks,'' \emph{J. Acoust. Soc. Am.}, vol. 153, no.~5, p.
  2690–2705, May 2023.

\bibitem{quazi81}
A.~Quazi, ``An overview on the time delay estimate in active and passive
  systems for target localization,'' \emph{IEEE Trans. Acoust., Speech, and
  Signal Process.}, vol.~29, no.~3, pp. 527--533, Jun. 1981.

\bibitem{huang01}
Y.~Huang, J.~Benesty, G.~W. Elko, and R.~M. Mersereati, ``{Real-time passive
  source localization: A practical linear-correction least-squares approach},''
  \emph{IEEE Trans. Speech and Audio Process.}, vol.~9, no.~8, pp. 943--956,
  Nov. 2001.

\bibitem{TesMeyBee:J20}
A.~Tesei, F.~Meyer, and R.~Been, ``Tracking of multiple surface vessels based
  on passive acoustic underwater arrays,'' \emph{J. Acoust. Soc. Am.}, vol.
  147, no.~2, pp. EL87--EL92, Feb. 2020.

\bibitem{ZhaMey:J24}
W.~Zhang and F.~Meyer, ``Multisensor multiobject tracking with improved
  sampling efficiency,'' \emph{IEEE Trans. Signal Process.}, vol.~72, pp.
  2036--2053, 2024.

\bibitem{MeyKroWilLauHlaBraWin:J18}
F.~Meyer, T.~Kropfreiter, J.~L. Williams, R.~A. Lau, F.~Hlawatsch, P.~Braca,
  and M.~Z. Win, ``Message passing algorithms for scalable multitarget
  tracking,'' \emph{Proc. {IEEE}}, vol. 106, no.~2, pp. 221--259, Feb. 2018.

\bibitem{MeyTesWin:C17}
F.~Meyer, A.~Tesei, and M.~Z. Win, ``Localization of multiple sources using
  time-difference of arrival measurements,'' in \emph{Proc. IEEE ICASSP-17},
  Mar. 2017, pp. 3151--3155.

\bibitem{gustafsson03tdoa}
F.~Gustafsson and F.~Gunnarsson, ``Positioning using time-difference of arrival
  measurements,'' in \emph{Proc. IEEE ICASSP-03}, vol.~6, Hong Kong, China,
  Apr. 2003, pp. 553--556.

\bibitem{WilLau:J14}
J.~L. Williams and R.~Lau, ``Approximate evaluation of marginal association
  probabilities with belief propagation,'' \emph{{IEEE} Trans. Aerosp.
  Electron. Syst.}, vol.~50, no.~4, pp. 2942--2959, Oct. 2014.

\bibitem{MeyBraWilHla:J17}
F.~Meyer, P.~Braca, P.~Willett, and F.~Hlawatsch, ``{A scalable algorithm for
  tracking an unknown number of targets using multiple sensors},'' \emph{{IEEE}
  Trans. Signal Process.}, vol.~65, no.~13, pp. 3478--3493, Jul. 2017.

\bibitem{barShalom11}
Y.~Bar-Shalom and X.~T. P.~K. Willet~and, \emph{{Tracking and Data Fusion: A
  Handbook of Algorithms}}.\hskip 1em plus 0.5em minus 0.4em\relax Storrs, CT,
  USA: Yaakov Bar-Shalom, 2011.

\bibitem{reid79}
D.~B. Reid, ``An algorithm for tracking multiple targets,'' \emph{IEEE Trans.
  Autom. Control}, vol.~24, no.~6, pp. 843--854, Dec. 1979.

\bibitem{AruMasGorCla:02}
M.~S. Arulampalam, S.~Maskell, N.~Gordon, and T.~Clapp, ``{A tutorial on
  particle filters for online nonlinear/non-Gaussian {B}ayesian tracking},''
  \emph{{IEEE} Trans. Signal Process.}, vol.~50, no.~2, pp. 174--188, Feb.
  2002.

\bibitem{BicLiBen:B08}
P.~Bickel, B.~Li, and T.~Bengtsson, ``Sharp failure rates for the bootstrap
  particle filter in high dimensions,'' in \emph{Pushing the Limits of
  Contemporary Statistics: Contributions in Honor of Jayanta K. Ghosh},
  vol.~3.\hskip 1em plus 0.5em minus 0.4em\relax Beachwood, OH, USA: Inst.
  Math. Statist., 2008, pp. 318--329.

\bibitem{ZhaMey:21}
W.~Zhang and F.~Meyer, ``Graph-based multiobject tracking with embedded
  particle flow,'' in \emph{Proc. IEEE RadarConf-21}, Atlanta, GA, USA, May
  2021.

\bibitem{MerDouFre:00}
R.~van~der Merwe, A.~Doucet, N.~de~Freitas, and E.~Wan, ``The unscented
  particle filter,'' in \emph{Proc. NIPS-00}, Denver, CO, USA, Dec. 2000, pp.
  584--590.

\bibitem{DuaHua:07}
F.~Daum and J.~Huang, ``{Nonlinear filters with log-homotopy},'' in \emph{Proc.
  SPIE-07}, Aug. 2007, pp. 423--437.

\bibitem{DuaHua:09}
------, ``{Nonlinear filters with particle flow induced by log-homotopy},'' in
  \emph{Proc. SPIE-09}, May 2009, pp. 76--87.

\bibitem{MosChaCha:C16}
N.~{Moshtagh}, J.~{Chan}, and M.~{Chan}, ``Homotopy particle filter for
  ground-based tracking of satellites at {GEO},'' in \emph{Proc. AMOS-16},
  Maui, HI, Sep. 2016, pp. 1--6.

\bibitem{DuaHua:10}
F.~Daum, J.~Huang, and A.~Noushin, ``{Exact particle flow for nonlinear
  filters},'' in \emph{Proc. SPIE-10}, Apr. 2010, pp. 92--110.

\bibitem{DuaHua:13}
F.~Daum and J.~Huang, ``{Particle flow with non-zero diffusion for nonlinear
  filters},'' in \emph{Proc. SPIE-13}, May 2013, pp. 226--238.

\bibitem{DuaHuaNou:18}
F.~Daum, J.~Huang, and A.~Noushin, ``New theory and numerical results for
  {Gromov}'s method for stochastic particle flow filters,'' in \emph{Proc.
  FUSION-18}, 2018, pp. 108--115.

\bibitem{DaiDau:22}
L.~Dai and F.~Daum, ``On the design of stochastic particle flow filters,''
  \emph{IEEE Trans. Aerosp. Electron. Syst.}, vol.~59, no.~3, pp. 2439--2450,
  2023.

\bibitem{BunGod:16}
P.~Bunch and S.~Godsill, ``{Approximations of the optimal importance density
  using Gaussian particle flow importance sampling},'' \emph{J. Amer. Statist.
  Assoc.}, vol. 111, no. 514, pp. 748--762, Aug. 2016.

\bibitem{LiZhaoCoates:15}
Y.~Li, L.~Zhao, and M.~Coates, ``Particle flow auxiliary particle filter,'' in
  \emph{Proc. IEEE CAMSAP-15}, Cancun, Mexico, Dec. 2015, pp. 157--160.

\bibitem{PalCoa:19}
S.~Pal and M.~Coates, ``Particle flow particle filter using {Gromov's}
  method,'' in \emph{Proc. IEEE CAMSAP-19}, Guadeloupe, France, Dec. 2019, pp.
  634--638.

\bibitem{LiCoates:17}
Y.~Li and M.~Coates, ``Particle filtering with invertible particle flow,''
  \emph{IEEE Trans. Signal Process.}, vol.~65, no.~15, pp. 4102--4116, Aug.
  2017.

\bibitem{khan2017bayesian}
M.~A. Khan, A.~De~Freitas, L.~Mihaylova, M.~Ulmke, and W.~Koch, ``Bayesian
  processing of big data using log homotopy based particle flow filters,'' in
  \emph{Proc. IEEE SDF-17}, 2017.

\bibitem{li2017sequential}
Y.~Li and M.~Coates, ``Sequential {MCMC} with invertible particle flow,'' in
  \emph{Proc. IEEE ICASSP-17}, 2017, pp. 3844--3848.

\bibitem{BarWilTia:B11}
Y.~Bar-Shalom, P.~K. Willett, and X.~Tian, \emph{{Tracking and Data Fusion: A
  Handbook of Algorithms}}.\hskip 1em plus 0.5em minus 0.4em\relax Storrs, CT:
  Yaakov Bar-Shalom, 2011.

\bibitem{Mah:B07}
R.~Mahler, \emph{{Statistical Multisource-Multitarget Information
  Fusion}}.\hskip 1em plus 0.5em minus 0.4em\relax Norwood, MA: Artech House,
  2007.

\bibitem{MeyWil:J21}
F.~Meyer and J.~L. Williams, ``Scalable detection and tracking of geometric
  extended objects,'' \emph{IEEE Trans. Signal Process.}, vol.~69, pp.
  6283--6298, 2021.

\bibitem{Poo:B94}
H.~V. Poor, \emph{An Introduction to Signal Detection and Estimation},
  2nd~ed.\hskip 1em plus 0.5em minus 0.4em\relax New York: Springer-Verlag,
  1994.

\bibitem{Cro-19}
D.~F. Crouse, ``Particle flow filters: biases and bias avoidance,'' in
  \emph{Proc. FUSION-19}, Ottawa, Canada, July 2019, pp. 1--8.

\bibitem{SchVoVo:J08}
D.~Schuhmacher, B.-T. Vo, and B.-N. Vo, ``A consistent metric for performance
  evaluation of multi-object filters,'' \emph{IEEE Trans. Signal Process.},
  vol.~56, no.~8, pp. 3447--3457, 2008.

\bibitem{Tuk:B77}
J.~Tukey, \emph{Exploratory Data Analysis}.\hskip 1em plus 0.5em minus
  0.4em\relax Addison-Wesley, 1977.

\bibitem{KocVan:J97}
W.~Koch and G.~Van~Keuk, ``Multiple hypothesis track maintenance with possibly
  unresolved measurements,'' \emph{{IEEE} Trans. Aerosp. Electron. Syst.},
  vol.~33, no.~3, pp. 883--892, 1997.

\bibitem{LiaKroMey:J23}
M.~Liang, T.~Kropfreiter, and F.~Meyer, ``A {BP} method for
  track-before-detect,'' \emph{IEEE Signal Process. Lett.}, vol.~30, pp.
  1137--1141, 2023.

\bibitem{LiaMey:J24}
M.~Liang and F.~Meyer, ``Neural enhanced belief propagation for multiobject
  tracking,'' \emph{IEEE Trans. Signal Process.}, vol.~72, pp. 15--30, 2024.

\end{thebibliography}




\end{document}